\documentclass{aa} 

\usepackage{graphicx}

\usepackage[utf8]{inputenc}
\usepackage{natbib}
\usepackage{makeidx}
\usepackage{multirow}
\usepackage{multicol}
\usepackage[dvipsnames,svgnames,table]{xcolor}
\usepackage{caption}
\usepackage{subcaption}
\usepackage{epstopdf}
\usepackage{ulem}
\usepackage{hyperref}
\usepackage{amsmath}
\usepackage{amssymb}
\usepackage[OT2,T1]{fontenc}
\hypersetup{colorlinks=true, citecolor=blue}
\DeclareSymbolFont{cyrletters}{OT2}{wncyr}{m}{n}
\DeclareMathSymbol{\Sha}{\mathalpha}{cyrletters}{"58}

\def\R{\mathbb{R}}
\def\Z{\mathbb{Z}}

\newcommand{\sinc}{\text{sinc}}

\def\rank{\mbox{rank}}

\newcommand{\T}{\mathsf{T}} 

  \newcommand{\G}{\mathcal{G}} 
  \newcommand{\FTG}{\widetilde{\mathcal{G}}} 
  \newcommand{\FTP}{\widetilde{\Pi}} 
  \newcommand{\Sl}{\mathbf{s}} 
  \newcommand{\FTS}{\widetilde{\mathbf{s}}} 
  \newcommand{\WF}{\boldsymbol{\varphi}} 
  \newcommand{\FTWF}{\widetilde{\boldsymbol{\varphi}}} 
  \newcommand{\Noise}{\boldsymbol{\eta}} 
  \newcommand{\FTN}{\widetilde{\boldsymbol{\eta}}} 

\usepackage[varg]{txfonts}

\begin{document} 

\title{Super-resolution wavefront reconstruction}
\subtitle{}
\titlerunning{Super-resolution wavefront reconstruction}
\authorrunning{S. Oberti et al.}

\author{Sylvain Oberti\inst{1}, Carlos Correia\inst{2}, Thierry Fusco\inst{3,4}, Benoit Neichel\inst{3} and Pierre Guiraud\inst{5}}

\institute{European Southern Observatory,Karl-Schwarzschild-Str. 2, 85748 Garching bei Muenchen, Germany \and Space ODT -- Optical Deblurring Technologies Unip Lda, Porto, Portugal \and
 Aix Marseille Univ, CNRS, CNES, LAM, Marseille, France\and 
DOTA, ONERA, Université Paris Saclay, F-91123 Palaiseau, France \and Instituto de Ingeniería Matemática and Centro de Investigación y Modelamiento de Fenómenos Aleatorios - Valparaíso.  Universidad de Valparaíso, Valparaíso, Chile}

   \date{Received May 4th, 2022; Accepted August 2nd, 2022}
\abstract{Cutting-edge ground based astronomical instruments are fed by Adaptive Optics (AO) systems that aim at providing high performance down to the visible wavelength domain on 10 m class telescopes and in the near infrared for the first generation instruments of the Extremely Large Telescopes (ELTs). In both cases the ratio between the telescope diameter $D$ and the coherence length or Fried parameter $r_0$, $D/r_0$ the parameter defining the required number of degrees of freedom, becomes large. Thus, $D/r_0$ drives the requirement to reconstruct the incoming wavefront with ever higher spatial resolution. In this context, Super-Resolution (SR) appears as a potential game changer. Indeed, SR promises to dramatically expand the amount of spatial frequencies that can be reconstructed from a set of lower resolution measurements of the wavefront.}{SR is a technique that seeks to upscale the resolution of a set of measured signals. SR retrieves higher-frequency signal content by combining multiple lower resolution sampled data sets.
SR is well known both in the temporal and spatial domains. It is widely used in imaging to reduce aliasing and enhance the resolution of coarsely sampled images.This paper applies the SR technique to the bi-dimensional wavefront reconstruction. In particular,we show how SR is intrinsically suited for tomographic multi WaveFront Sensor (WFS) AO systems revealing many of its advantages with minimal design effort.}{We provide a direct space and Fourier-optics description of the wavefront sensing operation and demonstrate how SR can be exploited through signal reconstruction, especially in the framework of Periodic Nonuniform Sampling. Both meta uniform and nonuniform sampling schemes are investigated. We show that, under some conditions, both sampling schemes enable a perfect reconstruction of band-limited signals. Then, the SR bi-dimensional model for a Shack-Hartmann (SH) WFS is provided and the characteristics of the sensitivity function are analyzed. The SR concept is then validated with numerical simulations of representative multi-WFS SH AO systems. Finally, we explore the extension of the method to Pyramid WFSs.}{Our results show that combining several WFS samples in a SR framework grants access to a larger number of modes than the native one offered by a single WFS -- and that despite the fixed sub-aperture size across samples. We show that the wavefront reconstruction achieved with four WFSs can be equivalent to a single WFS providing a twice larger sampling resolution (linear across the telescope aperture).
We also show that the associated noise propagation is not degraded under SR. Finally, we show that the concept can be extended to the signal produced by single Pyramid WFS, with its four re-imaged pupils serving as multiple non-redundant samples.}{SR applied to WaveFront Sensing and its Reconstruction (WFR) offers a new parameter space to explore as it decouples the size of the subaperture from the desired wavefront sampling resolution. By cutting short with old assumptions, new, more flexible and better performing AO designs become now possible.} 


\maketitle
%
\newcommand{\ccc}[1]{{\color{orange}\textrm{#1}}}
\newcommand{\bnc}[1]{{\color{red}\textrm{#1}}}
\newcommand{\tfc}[1]{{\color{blue}\textrm{#1}}}
\newcommand{\soc}[1]{{\color{green}\textrm{#1}}}
\newcommand{\pg}[1]{{\color{cyan}\textrm{#1}}}

\section{Introduction}

The term \textit{Super-Resolution} (SR) as is used here is a technique found under the computational imaging field whereby the joint application of optical design and signal processing techniques is engineered to obtain higher-resolution (HR) data products from multiple low-resolution (LR) samples. SR techniques
are relatively standard in vision science whenever imagery data is available with sub-pixel shifts between LR samples. First work on image upscaling was published in 1984 [\cite{Tsai1984}] and the term "Super-resolution" itself appeared around 1990 [\cite{Irani90}].
[\cite{Park03}] provides an excellent review, now roughly two decades old -- a sign of how well established SR imaging is at the time of writing. 

SR techniques are not new in the astronomical observation and instrumentation communities. A relevant example is the "drizzle" method developed to enhance the resolution of under-sampled HST images [\cite{Fruchter2002}] where the high spatial frequency information of the HST images are recovered  by combining sub-pixel dithered images.
In this paper we explore the application of SR to wavefront reconstruction for astronomical Adaptive Optics (AO) as proposed in the presentation [\cite{ObertiAO4ELT5}].

Reconstructing wavefronts instead of focal plane images is particularly interesting. Indeed, in the field of AO assisted astronomy instruments, the optical systems are typically designed to achieve a close to Nyquist sampling of the diffraction limited image. In other words, the images are well sampled and the potential to benefit from SR is moderate or null. On the other hand, the pupil image is typically not limited by diffraction in current AO systems. Concretely, the field stop is sufficiently opened to let the set of spatial frequencies of interest be sensed by the WFS. Moreover, the bi-dimensional wavefront surface can be described by a theoretically unlimited set of spatial frequencies, hence pushing the SR limit down to extremely small scales.

The reconstruction problem -- as it happens -- is often ill-conditioned because of insufficient LR data or is based on an ill-posed system model. Consequently,  some degree of regularisation is required to make it solvable, providing realistic and stable solutions. Even in the absence of multiple LR samples, the use of regularisation provides a means to estimate signal content beyond the native Nyquist-Shannon frequency. We call this strategy \textit{statistical SR} which is by now well-established in AO.  As a matter of fact, the use of regularised reconstruction in the form of an MMSE (Minimum-Mean-Square-Error) or MAP (Maximum-a-Posteriori) stochastic estimator [\cite{fusco01, conan14a, correia14}] or more generally Tikhonov regularisation inspired thereupon, is now commonplace. 
SR can also be approached \textit{geometrically} which we aim to expand on here. This is based on the idea that the samples from multiple wavefront samples, are combined to allow the reconstruction of higher-resolution spatial frequencies than what could be achieved by each single sample alone. Multiple sampled wavefront measurements (mostly across space
but we could similarly use video-sequences across time or a mix of both) are combined from all available
lower-resolution WFS measurements
provided each contains some form of unique phase information. 
In practice however, each SR facet can only be interpreted asymptotically: the \textit{geometric} when we approach the limit of least-squares reconstruction; the \textit{statistical} when fewer and fewer samples are used.

At this stage the following observation is paramount: multiple frames are natively granted for SR in classically-designed tomographic systems, where multiple WFSs are used, each looking at a single guide-star on a distinctive line-of-sight. The multitude of diverse WFSs back-projections at different ranges along those lines-of-sight, sample the turbulence in altitude at offset locations (grid points), which is sufficient to perform SR at every altitude range. The exception is the pupil plane where all sampling grids perfectly overlap making the relative phase information vanish. Depending on the geometric configuration, this may be so as well at definite altitudes for which the relative grid offsets are integer multiples of one sampling step.
\\

Building on the footsteps of [\cite{ObertiAO4ELT5}], the SR concept has been implemented and investigated for several AO applications.
For instance, the shape of a high order DM has been estimated by reconstructing a posteriori a temporal sequence of LR measurements from a single SH WFS, with proper fractional subaperture shifts applied to each temporal sample [\cite{Woillez2019}].

But the main applications of SR lies in tomographic AO systems where the signal from multiple WFS are combined.
A first prominent example include [\cite{ellerbroek01, wang12}] who tomographically reconstructed the 3D wavefront with an over-sampling factor of up to 2 from many identical WFS conjugated to the pupil-plane. Yet the principles of SR (\textit{statistical} SR in this case) were not called upon at that time.

In [\cite{ObertiAO4ELT5}], numerical simulations described the application of SR to GALACSI NFM [\cite{Oberti2018}], the Laser Tomography AO (LTAO) mode in operation at the Very Large Telescope (VLT). In fact, SR is already built-in the tomographic reconstruction process. Indeed, the 4 WFSs exhibit slight relative misregistrations that are not compensated by hardware but calibrated and taken into account in the system model, the tomographic Interaction Matrix, hence enabling SR through the MMSE reconstruction process.
More recently, MAORY the ELT Multi Conjugate Adaptive Optics (MCAO) system, has incorporated SR in its design. The concept has been analyzed via end-to-end simulations, with first studies enabling SR in altitude layers without adding misregistrations in the pupil plane [\cite{Oberti2019, Busoni2019}], and later taking advantage of the full \textit{geometrical} SR by introducing voluntary shifts and rotations between the sampling grids [\cite{Agapito2020}]. With such misregistrations - or diversity of sampling geometries - super-resolved wavefronts can be reconstructed in all layers, including the DM plane. As a result, it becomes possible to choose to either gain in aliasing reduction at iso number of subapertures or reduce the number of subapertures while preserving a similar level of performance. In this case, SR opens the door to new possibilities for WFS design choices, that allow increasing the subaperture size and fitting more pixels per subaperture, for example to deal with LGS spot elongation.
Consistent results have been obtained for several other tomographic AO systems, as for instance the HARMONI Laser Tomography Adaptive Optics (LTAO) mode [\cite{2022JATIS}] the Giant Magellan Telescope (GMT) GLAO and LTAO systems [Van Dam 2017, priv. comm.], the future VLT MCAO system - MAVIS - [\cite{Cranney2021},\cite{MAVIS2020}], and KAPA, the Keck Observatory LTAO system [\cite{wizinowich20}].
\\

\begin{figure*}[h]
	\begin{center}
            \includegraphics[width=0.5\textwidth,
             angle=0]{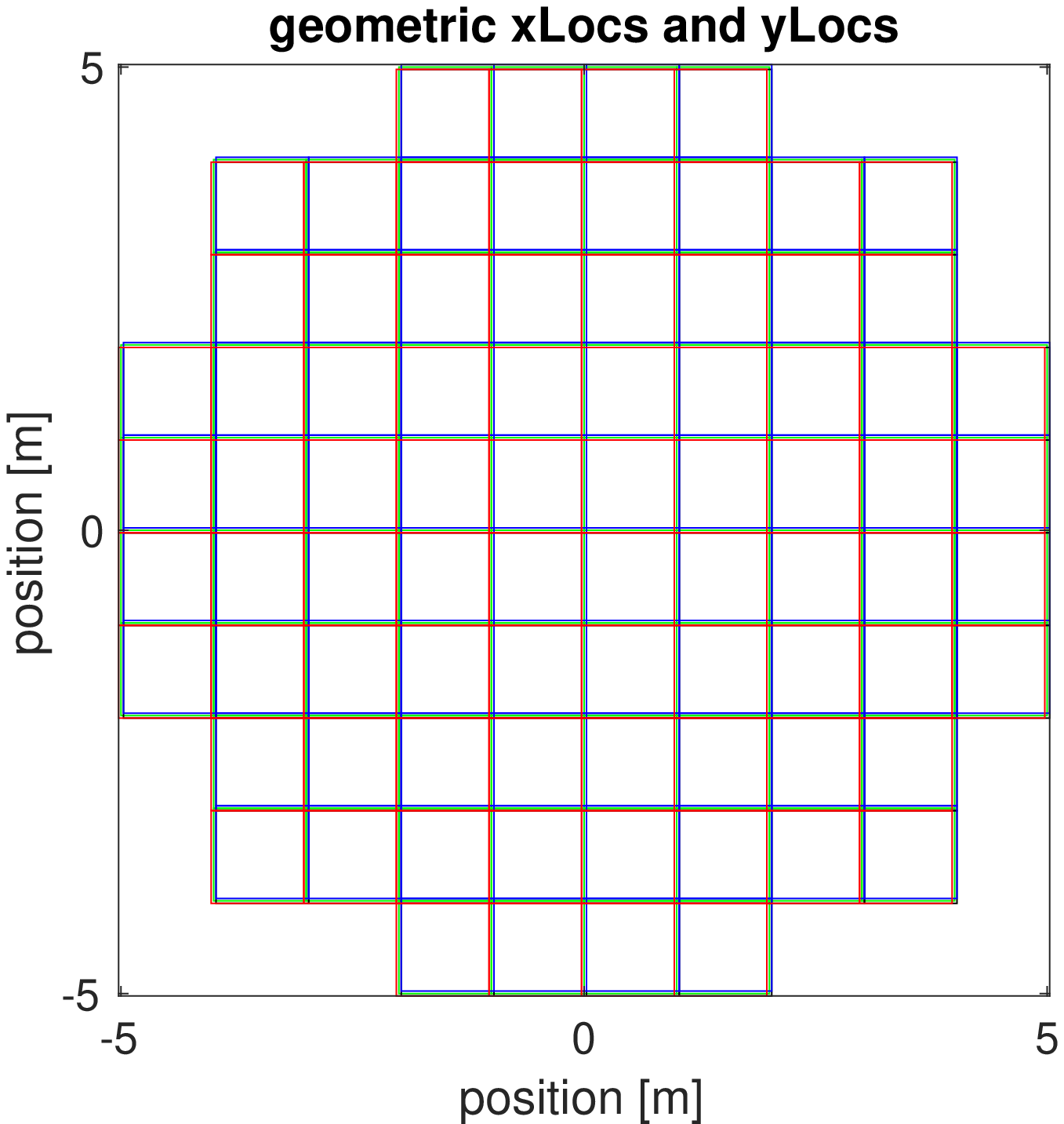}\includegraphics[width=0.5\textwidth,
             angle=0]{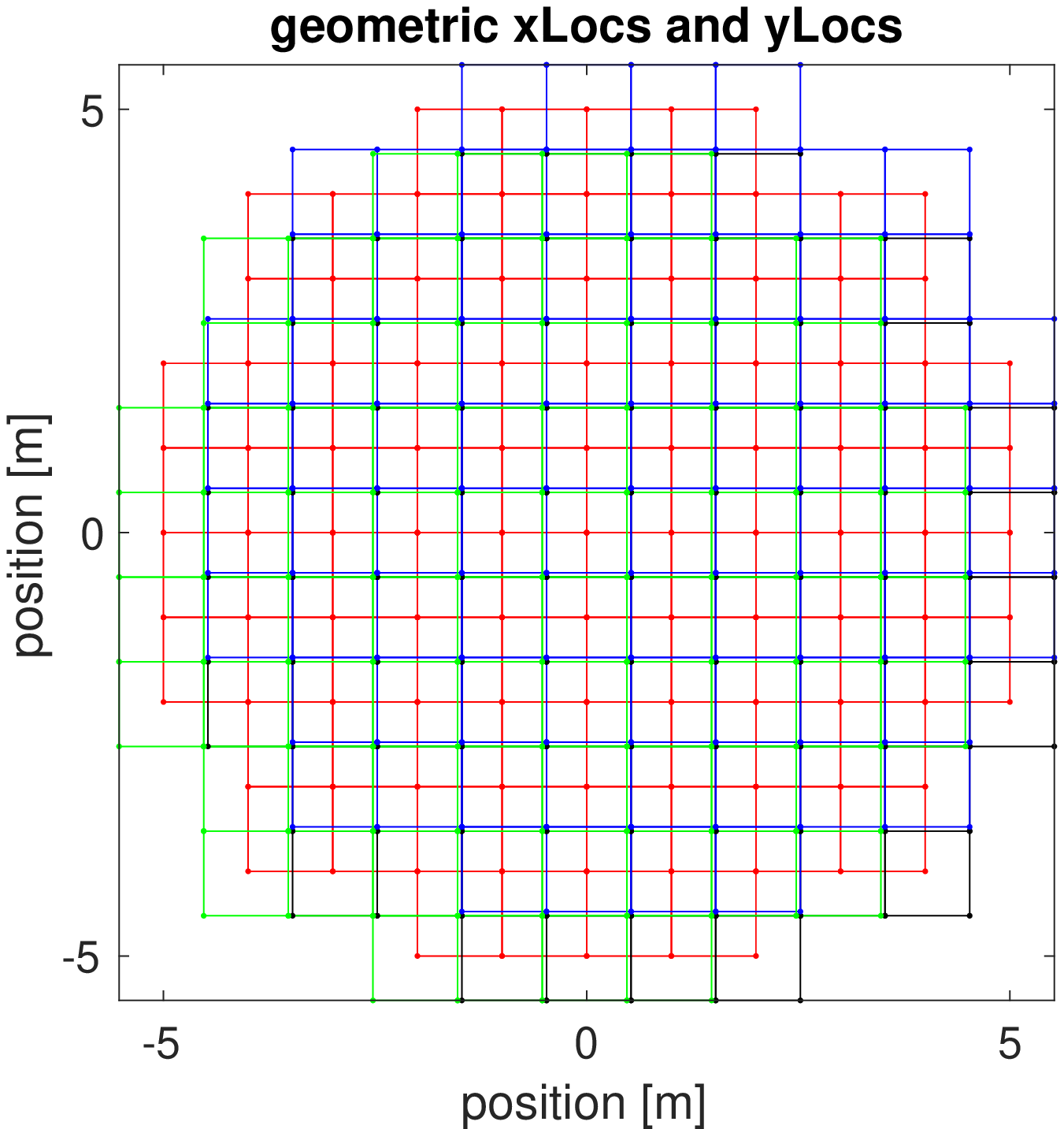}
	\end{center}
	\caption[]
	{\label{fig:alignedAtGround} (Left:) Aligned lenslets illustration. (Right:) Offset lenslet grids by 1/4 of a sub-aperture. The four colours used (red, green, blue and black) no longer superimpose.
 }
\end{figure*}

A meta uniform sampling can be simply achieved with 4 sampling grids by introducing proper offsets of half a subaperture width as depicted in Fig. (\ref{fig:alignedAtGround}). It appears clearly that with grids of finite size, the uniformity is not granted at the edge of the aperture. 
To enable SR at the ground but also over most conjugation altitudes, one obvious way is to break sampling and reconstruction symmetries (or regularities), and diversify the measurements (i.e. avoid redundancy) by adopting a nonuniform sampling. The most straightforward way is to offset and rotate the sampling grids with respect to one another. 
Other non uniform sampling options can be envisioned. Some examples are presented in Appendix \ref{AppB}. 

SR promises to support the control a $N_\text{act}\times N_\text{act}$ degree-of-freedom Deformable Mirror (DM) from measurements of several WFS each of a number of sampling elements $N_\text{sa} \times N_\text{sa}$ much smaller than $N_ \text{act}$, typically $N_\text{sa} \approx  \{1/2-2/3\} N_ \text{act}$ depending on the number of WFS. 

SR greatly relaxes the requirements on WFS alignment.  
For our application domain, we break free from registering the DM to the WFS as \textit{accurately} as possible, yet once done we need to know the registration \textit{precisely} in order to forward model it and estimate higher-order information. In other words, SR allows us to trade accuracy for precision. Although this distinction may seem elusive it is actually fundamental.\\
Remark: The registration shall indeed be calibrated \textit{precisely}, but not more \textit{precisely} than what classical AO systems require. Indeed, [\cite{Agapito2020}] suggests that the super-resolved cases are robust to model calibration errors: the performance is less sensitive to an error in the knowledge of the introduced misregistration amplitude than that of the classical MCAO system with co-aligned lenslet array grids in the pupil plane.

Furthermore, for multi-WFS AO systems, asymmetric sampling may reduce the unseen or badly-seen modes [\cite{neichel09}].
\\
It does not go unnoticed the commonalities with other phase diverse techniques in imaging and AO. Whereas classical phase-diversity relies on introducing a longitudinal phase encoding in the wavefront plane {\it along} the propagation path [\cite{gerchberg74}], the technique we explore here introduces it transversely with respect to the propagation. 
SR operated on fused data from all available lower-resolution WFS measurements provided each contains some form of unique phase information -- the basic prerequisite for super resolution. The unique phase information can be used to separate the aliased high-frequency from the content in low-frequencies of interest, and the higher-resolution wavefront can be accurately reconstructed.

The multiple samples can be acquired sequentially following the same principle as long as the observed scene (the wavefront in our case) remains static. 

\section{Paper outline}

In the following, the SR technique is first described from a theoretical point of view in the general and mono-dimensional case (\S \ref{PRINC}). In particular, we highlight the provided benefits in terms of sensitivity to higher spatial frequencies and aliasing reduction. We show that the signal can be reconstructed without aliasing with uniform or nonuniform samples (under some conditions) at the Nyquist-Shannon average sampling rate. Then, the bi-dimensional case is investigated further through analytical models (\S \ref{BiMod}). For SH WFSs, the sensitivity functions (\S \ref{SNR}) are studied in the diffraction limited and extended source cases. In \S \ref{NUMMODEL}, the application of SR to multi WFS AO systems is then analysed via numerical models, by evaluating the reconstruction error and the noise propagation.
Finally, we propose an extension of the SR concept to a single WFS, in the specific case of the Pyramid WFS (\S \ref{PWFS}).

\section{Super-resolution principle}\label{PRINC}

In this section, for the sake of clarity, we will first recall the super-resolution principle in the case of a mono-dimensional signal sampled by multiple periodic grids. Then, the main results applicable to AO will be highlighted, in the direct space and the Fourier domain.
This chapter provides a general description in the sense that the described SR technique can be applied to any continuous signal. In the specific case of adaptive optics, it may be applied to any kind of WFS, measuring the phase, its derivative, its laplacian or another linear transformation of the phase.

\subsection{Super-resolution framework in the direct space}

\subsubsection{Combination of several sampling grids: general case}
Let $u:=\{u_n\}_{n\in\Z}$ be a sequence of $\R$ and $f:\R \to \R$ a real function. If $f$ is discretely sampled at the points of  the sequence $u$, then the result of the sampling is represented by the distribution
\begin{equation}\label{DEFSAMPL}
f_u(x):=\sum_{n=-\infty}^{+\infty}f(x)\delta(x-u_n) \qquad\forall\,x\in\R.
\end{equation}

It follows from Eq. \eqref{DEFSAMPL},  that if we own $N$ sampling grids $u^1,u^2,\dots, u^{N}$, and that if $w$ is any sequence whose set of points is the union of the points of $u^1,u^2,\dots, u^N$, then 
\begin{equation}\label{SAMSUM}
f_w(x)=\sum_{\ell=1}^Nf_{u^\ell}(x)\qquad\forall\,x\in\R.
\end{equation}
In other words, the sum of $N$ sampling of $f$ at the points of the sequences $u^1,u^2,\dots, u^N$ is a sampling of $f$ at the union of the sequences that is $f_w$. 
For the AO scientist, each sampling sequence stands for the sampling grid corresponding to one WFS, composed of sampling points in the wavefront space, basically located of the center of each subaperture. In the following, without loss of generality and for any considered dimension, we will use the denomination \textit{sampling grid} instead of \textit{sampling sequence}. The quantity $N$ stands for the number of WFS samples at play, i.e. the number of WFSs or the number of temporal wavefront measurements that are combined. The sampling cell -- that we call a \textit{subaperture} -- is nominally the squared subaperture that averages the wavefront within its bounds. 

\subsubsection{Periodic Nonuniform Sampling}

Having quickly introduced the general framework, let us now consider a case of particular interest for the AO community: the Periodic Nonuniform Sampling (PNS) case. The PNS consists in combining several uniform sampling grids into a single nonuniform meta-sampling grid. The PNS is a scheme of deterministic sampling that was first introduced by [\cite{Kohlenberg1953}].

Let us consider $N$ uniform sampling grids $u^1,u^2,\dots, u^{N}$ with the same sampling step $h$. In AO, this corresponds to a set of $N$ WFSs with $h$ the spatial distance between the centers of two neighbouring subapertures. The sampling step $h$ is classically equal to the subaperture width.
The first sampling grid $u^1$  is given by 
\[
u_n^1=nh + u^1_0\qquad\forall\, n\in\Z,
\]
where $u^1_0\in\R$ is the reference origin, that can be for instance the center of the telescope pupil. For any $\ell\in\{1,\dots,N\}$, the sampling grid $u^\ell$ is a translated 
of $u^1$ of a quantity $\Delta_\ell$. For the AO scientist, $\Delta_\ell$ is equivalent to a shift of the WFS sampling grid number $\ell$ relative to an arbitrary reference WFS whose $\ell=1$. So we have $\Delta_1=0$ and will suppose $\Delta_\ell\in(0,h)$ if $\ell\neq 1$. Then, for every $\ell\in\{1,\dots,N\}$, we have
\[
u^\ell_n=nh+u^1_0+\Delta_\ell \qquad\forall\, n\in\Z.
\]
Let $w$ be any sequence whose points are the union of those of the sampling grids $u^1,u^2,\dots,u^N$.
For example, $w$ can be taken such that 
\begin{equation}\label{WK}
u_q^\ell=w_{qN+\ell-1} \qquad\forall\, q\in\Z \quad\text{and}\quad \ell\in\{1,\dots,N\}.
\end{equation}
In this way, if $\Delta_{1}<\Delta_{2}<\dots<\Delta_{N}$, the sequence $\{w_n\}_{n\in\Z}$ is sorted in increasing order (its points are labelled according to their spatial order) and if $n-n'=N$ then $w_n$ and $w_{n'}$ are two successive points of a same sampling grid $u^\ell$. 
It is interesting to note here that the operation described in  Eq. \eqref{SAMSUM} resembles a spatial reconstruction process, where the sampling points are re-arranged with respect to their spatial coordinates. In other words, the sampling sequences are not concatenated but organized according to the order of the sampling points. A simple toy example is easy to grasp: a sinusoid that aliases on each WFS, when samples from multiple WFSs are rearranged, is no longer seen to alias (depending on $N$, $\Delta_\ell$ and the actual frequency of the sinusoid).

\subsubsection{From fractional sample shifts to meta uniform sampling}
In the case where the shifts $\Delta_\ell$ are multiple of $\frac{h}{N}$, that is
\begin{equation}\label{FRAC}
\Delta_\ell=(\ell-1)\frac{h}{N}\quad\forall\, \ell\in\{1,2,\dots,N\},
\end{equation}
then Eq. \eqref{WK} writes as  
\[
w_{qN+\ell-1}=u^\ell_q=qh+u^1_0+(\ell-1)\frac{h}{N}
= \left(qN+\ell-1\right)\frac{h}{N}+u_0^1
\]
for every $q\in\Z$ and $\ell\in\{1,\dots,N\}$.
Therefore, 
\[
w_{n}=n\frac{h}{N}+u_0^1 \qquad\forall\, n\in\Z. 
\]
Then,
\begin{equation}\label{COMB}
\sum_{\ell=0}^{N-1}f_{u^\ell}(x)=f_w(x)=\sum_{n=-\infty}^{+\infty}f(x)\delta\left(x-n\frac{h}{N}-u^1_0\right) =(f\cdot\Sha_{\frac{h}{N}})(x),
\end{equation}
where 
\[
\Sha_{\frac{h}{N}}(x):=\sum_{n=-\infty}^{+\infty}\delta\left(x-\frac{nh}{N}-u_0^1\right)
\]
is the Dirac comb of period $h/N$ centered at $u_0^1$.

Let us now consider a superposition of $N$ sampling grids composed of subapertures of size $h$. In each subaperture, the function $f=\bar{\Sl}_h$ is the result of averaging the signal $\Sl$ by means of a convolution with the top hat function $\chi_h:\R \to\R$ that is the uniform density over the interval $I_h:=[-h/2,h/2]$,

According to Eq. \eqref{COMB}, the signal sampled by the superposition of sampling grids, that we call \textit{meta uniform sampling}, is thus given by the distribution: 
\begin{equation}\label{GenDirect}
{\Sl}_{N,h}(x)=(\bar{\Sl}_h\cdot\Sha_{\frac{h}{N}})(x)=(\Sl\ast\chi_h)\cdot\Sha_{\frac{h}{N}}(x)\qquad\forall x\in\R.
\end{equation}

 Remark: Eq. (\ref{GenDirect}) suggests that the we can decouple the size
 of the subaperture $h$ from the effective sampling step $h/N$, i.e. generate a \textit{meta-sampler} with regular sampling points located every $h/N$ and whose subapertures have a size that is different from the sampling step, unlike in classical AO designs where the spatial sampling step and the subaperture width are the same.

\subsection{Mono-dimensional representation in the Fourier domain}

The Fourier transform of Eq. \eqref{GenDirect}  writes:

\begin{equation}\label{GenFourier}
    \hat{\Sl}_{N,h}(k)=h\hat{\Sl}(k)\sinc(\pi hk)\ast\Sha_{\frac{N}{h}}(k)
\end{equation}
with $k$ the spatial frequency in $m^{-1}$ when considering the wavefront sensing related spatial sampling case.
The Fourier transform of the function $\Sl$ is multiplied by the $\sinc$ function that depends only on the subaperture size $h$ and convolved by $\Sha_{\frac{N}{h}}$ a Dirac comb of period $N/h$. The latter convolution leads to replicating the spectrum in the frequency domain due to the periodic spatial sampling. The $\sinc$ function can be understood as a smoothing function, standing for the individual subaperture transfer function. There is no loss of information due to the $\sinc$ function except at the null points located at multiple integers of $N/h$ in the Fourier space.

Let us also define the super-resolution factor $\mathcal{F}$ as the ratio between the maximum spatial frequency $k_{\max}$ that can be reconstructed without aliasing and the native Nyquist frequency $\nu=1/2h$, that is
\begin{equation}\label{SR_Factor}
\mathcal{F}=\frac{k_{\max}}{\nu}.
\end{equation}

So, $\mathcal{F}$ stands for the gain in resolution achieved with $N$ sampling grids, with respect to the case of a single grid with sampling step $h$.

\medskip

The meta uniform sampling case (Eq. \ref{GenFourier}) carries the following key features:
\begin{itemize}
    \item From the Nyquist-Shannon Theorem we know that in the case of Eq. \eqref{GenFourier} the signal can be reconstructed up to a frequency of $k_{\max}=N/2h$  without aliasing. The super-resolution factor $\mathcal{F}$ is therefore equal to $N$. Thus, combining properly shifted sampling grids leads to increasing the spatial resolution by a factor $\mathcal{F}$ hence decreasing the aliasing.
    \item The subaperture size is preserved hence the sensitivity function remains the same as the native one.
    \item SR offers the capability to design a \textit{meta-sensor} whose sampling step is independent from the subaperture size. Such independence allows to decouple the desired spatial sampling period from the sensitivity function in multi WFS AO systems, that is a new design paradigm.
\end{itemize}

Considering a super-resolution factor $\mathcal{F}$ of 2, it is interesting to note that there is no overlap of the spectra anymore until the first null of the $\sinc$ function. The illustration hereafter highlights the benefit of such feature in terms of aliasing reduction. 

\begin{figure}[ht!]
\centering
\includegraphics[scale=0.5]{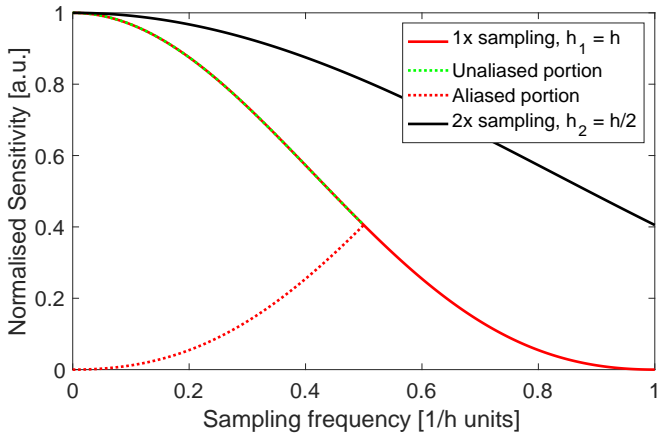}
\caption{Normalised sensitivity of a virtual \textit{averaging-and-sampler} sensor used for
  illustration purposes only. The black curve corresponds to the sensitivity function of a sampling grid of step $h/2$. The dotted curve represents the sensitivity of a sampling grid of step h. The red curve is obtained by dealiasing the spectral sensitivity up to $1/h$, i.e. the first null of the $\sinc$ function with a super-resolution factor of 2, that is to say by combining two sampling grids of step h relatively shifted by $h/2$ with respect to each other.}
\label{fig:aliasing2}
\end{figure}

The sensitivity is limited by the $\sinc$ function in the Fourier space (Eq. \eqref{GenFourier})
and by the aliasing of its replicated self. On Figure (\ref{fig:aliasing2}) we can see that beating the $\nu=1/2h$ spatial resolution limit leads to aliasing reduction when introducing a SR factor $\mathcal{F}=2$ (red curve). However, the sensitivity is greatly reduced with respect to a case with a twice larger native sampling frequency (black curve). With a SR factor $\mathcal{F}>2$, it would be possible to go beyond a resolution of $2\nu=1/h$ (not shown on this figure). Nevertheless, the sensitivity would again be limited by the $\sinc$ function that depends only on the subaperture size. For spatial frequencies larger than the first null of the $\sinc$ function, the sensitivity (red curve) is much lower than what would be achieved with a four times larger native sampling. This illustration indicates that noise propagation shall be analysed with care in presence of super-resolution. Indeed, on the one hand the sensitivity is limited by the size of the subaperture. On the other hand, a larger subaperture size allows reaching a higher signal to noise ratio. These two opposite effects shall be taken into account to evaluate the noise propagation of a super-resolved system. 

\subsection{Signal reconstruction: number of sampling grids and optimal offsets}
In this section, we wish to address signal reconstruction by analyzing the number of sampling grids required for a reconstruction without aliasing, as a function of their relative spatial offsets.
From the Nyquist-Shannon sampling theorem,  we know that the choice of $N$ offsets according to Eq. \eqref{FRAC}, gives a meta uniform sampling providing a SR factor $\mathcal{F}=N$. We analyse here whether or not there are other ways to select the offsets, in order to reach a higher SR factor, or at least to preserve it, without adding extra sampling grids.\\

In Appendix \ref{APP}, we analyze the reconstruction of band-limited signals by convolutions with $\sinc$ interpolation functions, in the PNS framework.
Under conditions that simplify the analysis, but may be too restrictive, we show that with $N$ sampling grids, $\mathcal{F}$ cannot be higher than $N$, independently of the offsets choice. Moreover, the meta uniform sampling is the only solution with $N$ sampling grids that reaches $\mathcal{F}=N$, i.e. that allows to reconstruct the signal without aliasing until the frequency $N\nu$. With nonuniform configurations, i.e. with samples not taken equally spaced, more than $N$ sampling grids are necessary. Otherwise, the departure from the meta uniform sampling scheme will generate some reconstruction error. We can show that this error term is bounded and converges to zero when tending towards the meta uniform sampling. However, we will see hereafter that other hypotheses and more elaborated reconstruction approaches allow to show that some well chosen non uniform sampling schemes can reach the same performance as the meta uniform sampling.

In practise, it may be tricky to place the sampling grids exactly at the positions required to generate a meta uniform sampling. So, we wonder how many sampling grids are needed to achieve a given SR factor $\mathcal{F}$, even if the sampling grids are disposed without particular care, possibly with random offsets.  We show that with $N=2n_0-1$ randomly positioned sampling grids, where $n_0=\left\lceil\mathcal{F}\right\rceil$ is the ceiling function of $\mathcal{F}$, we can always reach a SR factor at least equal to $\mathcal{F}$.

To be specific and give some concrete examples, if we want to reconstruct a signal whose maximal frequency 
satisfies $\nu < k_{max} \leq 2\nu$, we need  $1 < \mathcal{F}\leq2$, and $n_0=2$. So, in the mono-dimensional case, we may use 2 sampling grids providing a meta uniform sampling, or 3 sampling grids randomly positioned, excluding the trivial case where some offsets are multiple of the native sampling step. In the bi-dimensional case, we can show that 4 sampling grids are needed with a meta uniform sampling scheme, while 9 randomly positioned sampling grids are sufficient.\\

As previously mentioned, with a more elaborated approach based on Lagrange polynomials, it is possible to reconstruct non-uniformly sampled signals without aliasing, even outside the PNS framework. Indeed, the Paley–Wiener–Levinson theorem ([\cite{Marvasti2021, Wiener30, PW34,Levinson1940}]), which generalizes the Whittaker–Shannon–Kotelnikov  sampling theorem [\cite{Marvasti2021}] from uniform to nonuniform samples, states that a band-limited signal can be reconstructed from its samples if the two following sufficient conditions are respected:
\begin{itemize}
\item According to [\cite{Beutler1966}], the average sampling rate $k_e$, that is the inverse of the average sampling step $h_e$, satisfies the generalized Nyquist-Shannon condition:
\begin{equation}\label{Nyquist-Shannon}
k_e=\frac{1}{h_e}\geq 2k_{\max}
\end{equation}
In the PNS framework, $h_e$ is always equal to $h/N$. To reach a SR factor $\mathcal{F}=N$, at least $N$ sampling grids are necessary.\\
\item The Kadec condition [\cite{Kad64}] states that the nonuniform sampling shall be constructed by relocating the points of a uniform sampling of step ${1}/{2k_{max}}$ no further than ${1}/{8k_{max}}$ from their original location.
\end{itemize}

We have seen that under some conditions, both uniform and nonuniform sampling enable a theoretically perfect reconstruction of the sampled band-limited signal. In the PNS framework, the $N$ sampling grids' offsets shall be close to but not necessarily equal to a multiple of $h/N$, in order to achieve optimal SR ($\mathcal{F}=N$). We will see in \S \ref{NUMMODEL} figure (\ref{fig:OLrec_WFE_vs_Shift}) that numerical simulations confirm the results from this section. The meta uniform sampling is indeed optimal in the sense that it provides the best reconstruction performance. Nevertheless, it is important to note that the performance is improved as soon as any non zero shift is introduced. Moreover, there is a shift range around the optimal shift that provides stable performance, which indicates its robustness. This shift range is consistent with the Kadec condition [\cite{Kad64}], for a perfect reconstruction with nonuniform sampling.

\section{Super-resolution Wavefront Reconstruction with Shack-Hartmann WFSs}

\subsection{Bi-dimensional model}\label{BiMod}
Let us now transpose the previous models to a SH-WFS [\cite{rigaut98}] whose measurements $\Sl(\mathbf{x})$ are well approximated  by the geometrical-optics linear model  
  \begin{equation}\label{eq:s_Gphi}
    \Sl(\mathbf{x}) = \G\WF(\mathbf{x}) + \Noise(\mathbf{x}),
  \end{equation}
  where $\G$ is a phase-to-slopes linear operator mapping
  aperture-plane guide-star
  wavefronts $\WF(\mathbf{x})$ onto WFS measurements over a bi-dimensional space indexed by $\mathbf{x} = [x,y]$;  $\Noise(\mathbf{x})$ represents white
  noise due to photon statistics,
  detector read noise and background photons. Both $\WF$ and $\Noise$ are zero-mean functions of Gaussian probability distributions and known covariance matrices $\Sigma_\phi$ and
  $\Sigma_\eta$ respectively. Noise is assumed both temporally and spatially uncorrelated. 
 
It can be shown for the nominal case when the subaperture size $d$ coincides with the sampling step $h$ [\cite{correia17}] that 
  \begin{equation}
    \G = \Sha\left(\frac{\mathbf{x}}{h}\right) \times \left[\Pi\left(\frac{\mathbf{x}-1/2}{d}\right) \otimes \nabla\right],
  \end{equation}
  where $\otimes$ is a 2-dimensional convolution product, $\times$ is a point-wise multiplication, $\Pi (\cdot)$ 
  is the top-hat separable function 

$\nabla$ is the gradient operator and $\Sha(\bf x)$ is a comb function (a bi-dimensional sum of
Dirac delta functions) that represents the sampling process. In the following, we will consider the classical case $d=h$.

Now the argument follows. If we properly combine four sets of measurements provided by convolutional operators 
\begin{equation}
\G = \left[
\begin{array}{c}
     \G_1  \\
     \G_2  \\
     \G_3  \\
     \G_4 
\end{array}
\right]
\end{equation}
with a relative
offset between them of $h/2$, corresponding to the meta uniform sampling in 2D, i.e.

\begin{eqnarray}\label{eq:meta2D}
\G_1  &=& \Sha\left(\frac{\mathbf{x}}{h}\right) \times \left[\Pi\left(\frac{\mathbf{x}-[1/2,1/2]}{h}\right) \otimes \nabla\right],\nonumber\\
\G_2 &=& \Sha\left(\frac{\mathbf{x}-[0,1/2]}{h}\right) \times \left[\Pi\left(\frac{\mathbf{x}-[1/2,1/2]}{h}\right) \otimes \nabla\right],\nonumber\\
\G_3 &=& \Sha\left(\frac{\mathbf{x}-[1/2,0]}{h}\right) \times \left[\Pi\left(\frac{\mathbf{x}-[1/2,1/2]}{h}\right) \otimes \nabla\right],\nonumber\\
\G_4 &=& \Sha\left(\frac{\mathbf{x}-[1/2,1/2]}{h}\right) \times \left[\Pi\left(\frac{\mathbf{x}-[1/2,1/2]}{h}\right) \otimes \nabla\right]
\end{eqnarray}
  it leads to a meta-SH system with half the sampling step and the same averaging cell size or subaperture width as in the native case
    \begin{equation}
    \bar\G = \Sha\left(\frac{\mathbf{x}}{h/2}\right) \times \left[\Pi\left(\frac{\mathbf{x}-1/2}{h}\right) \otimes \nabla\right],
  \end{equation}
therefore lifting the sub-aperture-imposed spatial-frequency cut-off
frequency which becomes in this case with SR factor $\mathcal{F}=2$
\begin{equation}\label{eq:f_sr}
 k_{max} = 2\nu = \frac{1}{h}
\end{equation}

Using the fact that the SH-WFS measurement model is a set of convolution integrals, treatment in the spatial-frequency domain is straightforward. 
  Let the Fourier-domain representation of Eq. (\ref{eq:s_Gphi})
  \begin{equation}\label{eq:S_FT}
    \FTS\left(\boldsymbol{\kappa}\right) = \FTG \FTWF\left(\boldsymbol{\kappa}\right) + \widetilde{\boldsymbol{\eta}}\left(\boldsymbol{\kappa}\right),
  \end{equation}

  with $\boldsymbol{\kappa} = (k_x,k_y) \in \mathbb{R}^2$ the frequency vector and symbol $\widetilde{\,\,\cdot\,\,}$ used for Fourier-transformed variables. The time-dependence is added in with recourse to the frozen-flow hypothesis. Using common transform pairs for the individual operations (see e.g.
  [\cite{oppenheim97}]), the Fourier
  representation 
  of the measurements in 
  Eq. (\ref{eq:s_Gphi}) given in 
 Eq. (\ref{eq:S_FT}) can be expanded to
    \begin{align} \label{eq:SH_WFS_meas_FT}
    \FTS\left(\boldsymbol{\kappa}\right) = &2 i \pi h \sum_{\mathbf{m}} \left(\boldsymbol{\kappa} + \mathbf{m}/\left(h/N\right) \right) \FTWF\left(\boldsymbol{\kappa} + \mathbf{m}/(h/N) \right)
     \nonumber\\
    &\hspace{30pt}\FTP\left(\boldsymbol{\kappa}h + \mathbf{m}\right)\times e^{i \pi (\boldsymbol{\kappa}h + \mathbf{m})}
    + \FTN\left(\boldsymbol{\kappa}\right) 
    \end{align}
 From Eq. (\ref{eq:SH_WFS_meas_FT}) it becomes apparent that the wavefront sensing operator $\FTG$ is not purely a spatial filter except for functions within the SH-WFS pass-band [\cite{ellerbroek05}]. 

\subsection{Trade-off sensitivity v. noise propagation}\label{SNR}

In practical terms, any interested AO practitioner may wonder quite righteously whether SR signal reconstruction above the nominal Nyquist-Shannon cut-off is associated to a somewhat larger noise propagation which could limit the prospective SR advantages. We will first address this key point in the spatial-frequency domain and later using the Karhunen-Loeve (KL) modal basis. We show in either case that SR grants comparable noise propagation without sacrificing effective sensitivity. 

\subsubsection{Sensitivity functions revisited: Fourier SNR}

We start by revisiting commonly-accepted expressions for sensitivity
functions, often called Fourier Signal-to-Noise Ratio (SNR) [\cite{verinaud04}].
The sensitivity defined as a
SNR (the ratio between
the produced signal and the noise affecting the measurement) is a
function of the effective number of photons and spot size on each
sub-aperture, as follows
\begin{equation}\label{eq:shwfs_sensitivity1}
SNR = \Xi^2_\text{SH}/\sigma^2_\text{ph}
\end{equation}
The SH-WFS sensitivity can be readily computed 
from Eq. \eqref{eq:SH_WFS_meas_FT} for functions in the pass-band assuming the compact form  [\cite{correia14}]
\begin{equation}\label{eq:shwfs_sensitivity2}
\Xi_\text{SH}^2 = \left(2i\pi h k_x \right)^2 \left(\frac{sin(\pi k_x
  )}{\pi k_x h}\right)^2 = \Xi_\text{grad}^2 \Xi_\text{aver}^2
\end{equation}
where we assume the effective SH signal to be the angle-of-arrival on each sub-aperture instead of the phase-difference at opposite sub-aperture edges (there is a normalisation factor $h$ between both). The 2D case is now straightforward to obtain. 

Let the following photon- and detector-noise expressions [\cite{roddier99, hardy98}]
\begin{equation} \label{eq:photon_noise}
\sigma^2_{\alpha,ph} = \frac{1}{8}\frac{1}{n_{ph}}\theta_{spot}^2
\end{equation}
\begin{equation}\label{eq:ron_noise}
\sigma^2_{\alpha,det} =   \frac{1}{12}\frac{\sigma_e^2}{n_{ph}^2}\left( \frac{N_s^2}{N_D}\right)^2 \theta_{spot}^2 
\end{equation}
with $n_{ph}$ the number of incident photons per frame and per sub-aperture,$\theta_{spot}$ the spot size, $\sigma_e$ the rms number of detector read-out photo-electrons, $N_s$ and $N_t$ are the window used for processing and sub-aperture diffraction sizes respectively. 
In the remainder we will restrict the analysis of photon-noise only for it is not far-fetched these days to consider practically zero-noise detectors. 

\begin{figure}[ht!]
\centering
\includegraphics[scale=0.5]{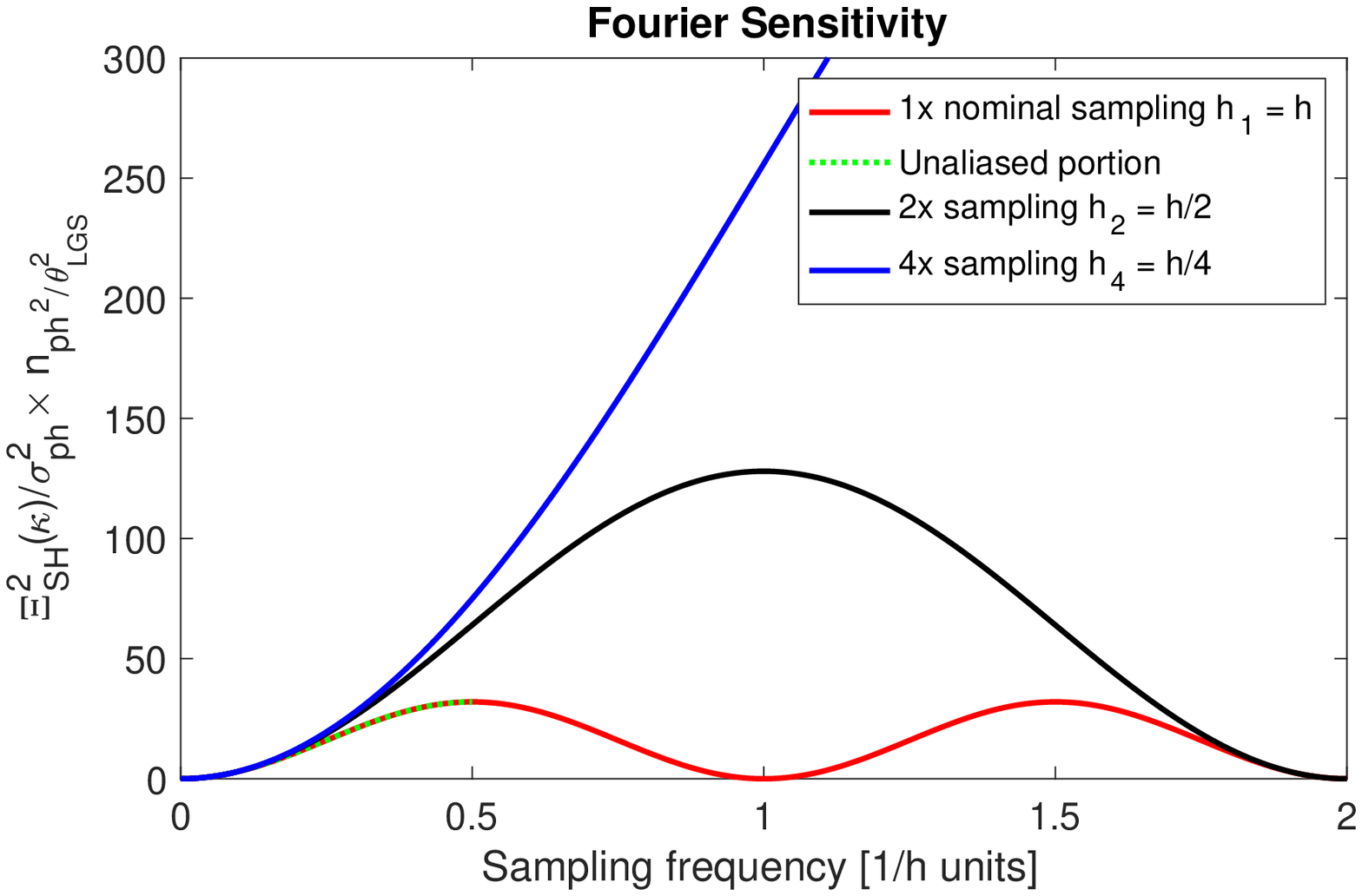} \includegraphics[scale=0.5]{Figures/fourierSNR_SH_seeingLimSpot_GridOff.eps}
\caption{Fourier SNR functions. Top: Fourier SNR for a diffraction-limited wavefront sensing case. (Eq. \eqref{GenDirect}). Bottom: Seeing-limited or laser spot sensing case. The number of incident photons is considered the same for all configurations. The bottom panel assumes a spot size with one unit.}
\label{fig:sensitivity}
\end{figure}

In diffraction-limited cases with point-sources, then $\theta_{spot} = \frac{\lambda}{h}$. It it straightforward to check that if one chooses to split the aperture into $n_{SA}$-times as many sub-apertures (linear across the pupil) the photon-noise is increased by a factor ${n^4_{SA}}$ resulting from the product of ${n^2_{SA}}$ less photons per sub-aperture and a diffraction spot size which is twice as large, resulting in another ${n^2_{SA}}$ factor for the bi-dimensional case. Considering all the measurements over the whole pupil, an average over ${n^2_{SA}}$ more measurement point is performed resulting in a final ${n^2_{SA}}$ sensitivity increase. This sensitivity increase is at the origin of the so-called "full-aperture gain" which leads to the use of a single full aperture spot measurement for tip and tilt modes. A comparative plot shown in the top panel of Fig. (\ref{fig:sensitivity}) illustrates this fact. This would lead to the conclusion that using a super-resolved meta SH-WFS setting leads to sensitivity gains over a twice-resolved SH and even to a 4x-resolved SH over the original sampling band $[0:\frac{1}{2h}]$ and some portions of the $[\frac{1}{2h}:\frac{1}{h}]$. Therefore, SR appears very well suited for AO applications with diffraction limited WFS, e.g. near infrared Natural Guide Star (NGS) based.

We do expect however to deploy multi-WFS mostly on laser-tomographic AO cases. In this case, splitting the aperture into less sub-apertures leads to small to none sensitivity gains because the spot is no longer a function of the diffraction but instead of a fixed size. From the bottom panel of Fig. (\ref{fig:sensitivity}) we can inspect the relative sensitivities: they are roughly the same until half the control radius of the nominal case $[0:\frac{1}{4h}]$ by which time the larger sub-aperture size averaging function (the sinus cardinal) starts bending down the sensitivity curve. In the SR case, we can only be as sensitive as the red curve, yet with the aliasing cut-off frequency a function of the over-sampling used. The periodic nature of the sensitivity function with an initial zero at $\kappa = 1/h\, m^{-1}$ tells us that an over-sampling of up to two is of practical relevance but not much beyond for a large noise amplification of frequencies with close-to-zero sensitivity is to be traded-off by increased sensitivity to frequencies beyond the native cut-off frequency.  

\begin{figure}[ht!]
\centering
\includegraphics[scale=0.5]{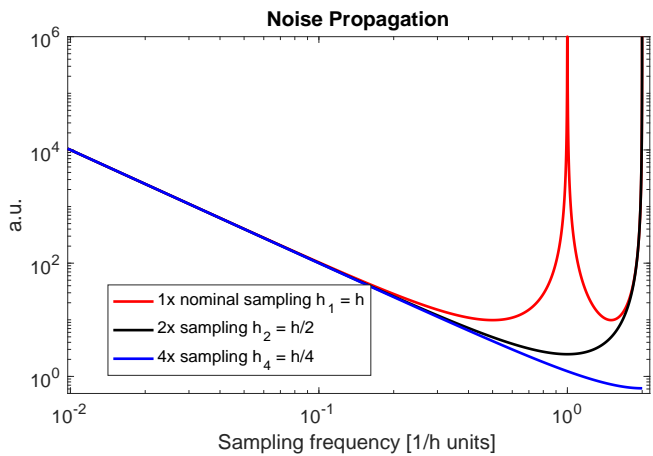} 
\caption{Noise propagation coefficient as a function of spatial frequency. This plots are computed as the inverse of the SNR functions from Fig. (\ref{fig:sensitivity}). The fluxes are scaled by the collecting area -- i.e. the square of the subaperture width. 
}
\label{fig:noisePropSpatFreqs}
\end{figure}

If we opt to evaluate the noise propagation instead we can see from Fig. (\ref{fig:noisePropSpatFreqs}) that SR will not lead to higher figures at low spatial-frequencies and will extend the effective maximum reconstructed frequency avoiding the zeros of the $\sinc^2$ function for which the noise propagation is infinite. 
The sensitivity evaluated on a single sub-aperture tells us little about how it is distributed over modes spanning the aperture. For tip-tilt it is easy to compute since the global estimate is a straight average of each sub-aperture's measurements, resulting in an effective $f^2$ gain (diffraction-limited sensing) or no gain (seeing or LGS spot limited sensing). In the next section we will study a more general case of modal reconstruction for multi-WFS AO systems.

\subsubsection{Noise propagation: modal analysis}

In order to properly treat the general case, the noise propagation will be considered on a per-mode basis, following [\cite{rigaut92}]
\begin{equation}\label{eq:noise_prop}
n.p. = diag \{R R^T \}
\end{equation}
where $R$ stands for the reconstructor matrix. We use this formulation instead of the original one provided by Eq. (14) in [\cite{rigaut92}] due to the use of regularised reconstruction. If the least-squares solution were adopted, we would find Rigaut's $\left(D^\T D\right)^{-1}$, with $D$ the system modal interaction matrix concatenating the sensor's response to each mode. 

Figure (\ref{fig:noisePropSpatFreqs}) -- which is essentially the inverse of Fig. (\ref{fig:sensitivity}) as it can readily be seen -- depicts the noise propagated as a function of spatial frequency. When accounting for the subaperture size in the computation of the noise propagation, we can confirm that SR configurations do not propagate more noise than classical configurations, along the lower spatial frequencies.

\subsection{Numerical application to multi WFS AO systems}\label{NUMMODEL}

In this chapter, we numerically simulate the SR concept applied to a multi SH-WFS AO system.
First of all, we can easily confirm that SR can be promoted by building a multi SH-WFS model and measuring the Interaction Matrix (IM) between the phase and the meta SH-WFS combining several LR WFS, in the configuration described in Table (\ref{tab:Configuration}).

\begin{table}[ht!]
\centering
\begin{tabular}{|c|c|}
\hline
     Telescope Diameter &   8\,m\\ 
    Number of WFSs & 4 SH-WFS \\ 
    Native resolution case & $n_{SA} = 40$ subapertures \\
    Lower resolution cases & $n_{SA} = 20$ or $10$ subapertures \\
    Turbulent wavefront & 1 layer in the pupil plane: KL modes\\
    \hline
\end{tabular}
\caption{Simulation configurations: the 4 Sh-WFS grids are either co-aligned as in a classical tomographic AO system or shifted with respect to each other. The super-resolution sampling scheme is shown in Fig. (\ref{fig:2D_Samp_Geom}) case (b) for a SR factor of 2, i.e. 4 WFS in 2D.}
\label{tab:Configuration}
\end{table}

\begin{figure}
     \centering
     \begin{subfigure}[b]{0.2\textwidth}
         \centering
         \includegraphics[scale=0.55]{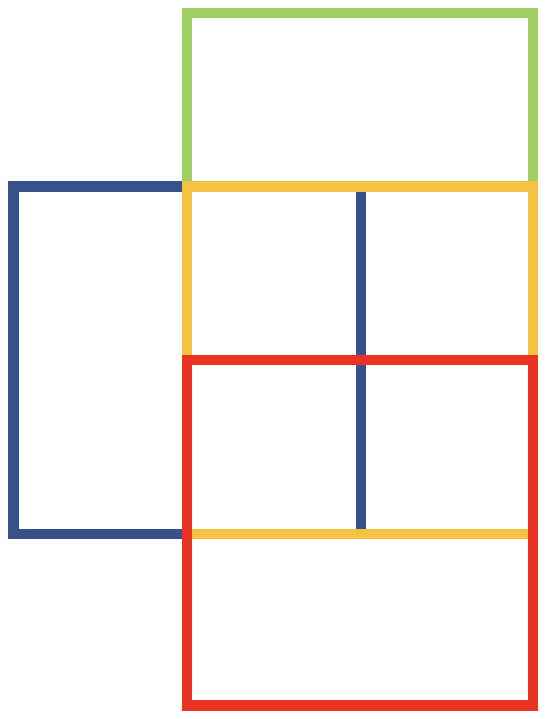}
         \caption{}
         \label{figleft}
     \end{subfigure}
     \begin{subfigure}[b]{0.2\textwidth}
         \centering
         \includegraphics[scale=0.55]{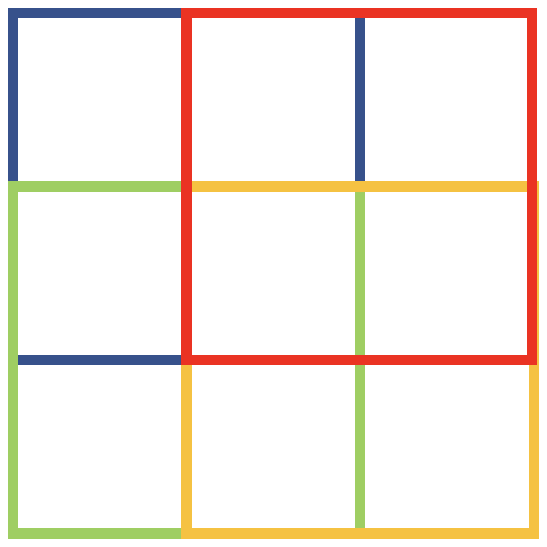}
         \caption{}
         \label{figright}
     \end{subfigure}
        \caption{Super-resolution configurations illustrated by showing the projection in the pupil plane of one subaperture from each WFS (1 color for each). a) The 2D surface is improperly paved with shifts along the sampling grid's axes. b) The 2D surface is properly paved with the addition of shifts at 45 degrees with respect to the sampling grid's axes, as in Eq. \eqref{eq:meta2D}.}\label{fig:2D_Samp_Geom}
\end{figure}

Let us compute a set of modal IMs, between the 1364 KL modes as input and the 4 SH-WFS slopes as output, this for all configurations (40x40, 20x20, 10x10, co-aligned or shifted).
The rank of the IM indicates the number of degrees of freedom that the super-resolved system is sensitive to. The Singular Value Decomposition (SVD) of the IM highlights the sensitivity of the system to the singular modes. In other words, the singular value associated to each singular mode indicates how well such mode is seen by the system.
\begin{figure}[ht!]
\begin{center}
    \includegraphics[scale=0.5]{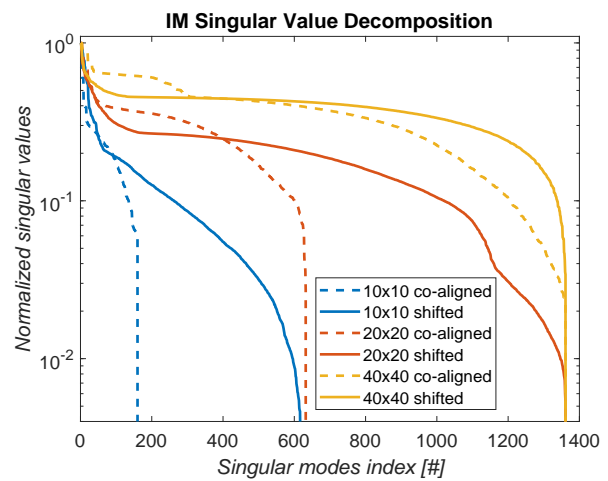} 
\end{center}
\caption{Singular Value Decomposition of a multi-WFS mono layer modal IM. The singular values are plotted as a function of the singular modes index. The solid lines corresponds to 4 co-aligned SH-WFS while the dashed lines represent the shifted configurations shown in Figure (\ref{fig:2D_Samp_Geom}) Case (b).}
\label{fig:IM_PCA}
\end{figure}
Figure (\ref{fig:IM_PCA}) highlights the benefit provided by the shifted configurations. In the classical cases (solid lines), the sensitivity drops to 0 for a KL index equal to the number of slopes measured by each individual SH-WFS. On the other hand with super-resolution (dashed lines), the sensitivity is enhanced and the sensitivity limit is pushed towards that of the co-aligned cases with double native resolution. Concretely, the super-resolved 10x10 (resp. 20x20) configuration provides some additional sensitivity up to the singular modes located at the sensitivity limit for the 20x20 (resp. 40x40) co-aligned configuration.

Let us now perform an open loop reconstruction of the phase by inverting the previously computed IMs in a MMSE fashion.
Note that each KL mode past the LR SH-WFS cut-off frequency
$\nu=1/(2h)$ aliases onto its corresponding low-frequency spectrum, yet the inversion of the meta SH-WFS IM plays the role of a numerical de-scrambler as if the samples were spatially reordered and therefore de-aliased.
\begin{figure}[ht!]
\begin{center}
\includegraphics[width=0.5\textwidth]{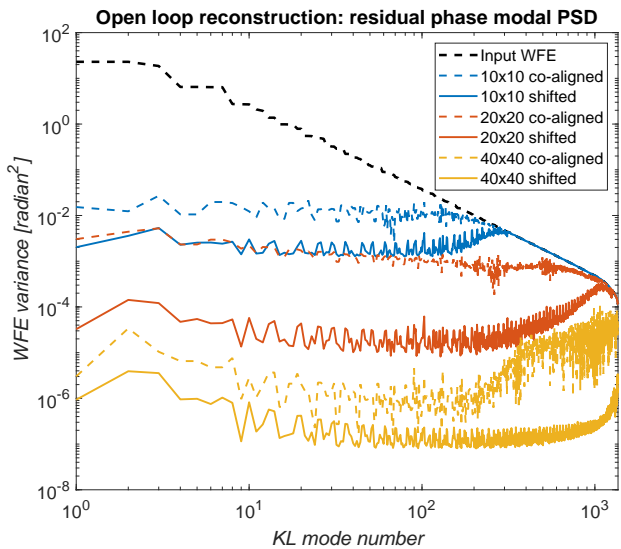} 
\end{center}
\caption{Open loop reconstruction of an incident phase corresponding to a Kolmogorov turbulence with $r_0=12.6$ cm. The dashed lines represent the co-aligned cases. The solid lines stand for the super-resolved cases achieved by combining 4 SH WFS properly shifted by half a subaperture. The same number of modes is reconstructed for the co-aligned cases and their respective shifted cases, that is up to 4.$\frac{\pi}{4}{n^2_{SA}}$ modes, a spatial resolution that would be provided by a single WFS with twice more subapertures per diameter}
\label{fig:OLrec_WFE_vs_SFreq}
\end{figure}
Figure (\ref{fig:OLrec_WFE_vs_SFreq}) indicates that the super-resolved configurations significantly reduce the reconstruction error with respect to the co-aligned cases. The cut-off frequency is twice higher than the respective classical case with as many reconstructed modes as number of sensing element per WFS $\frac{\pi}{4}{n^2_{SA}}$. Moreover, the aliasing features vanish close to the cut-off frequency for the super-resolved cases. Recent end-to-end closed loop simulations performed in the context of ELT tomographic AO systems (MAORY and HARMONI LTAO) confirm the benefit from SR in terms of aliasing reduction at iso number of subapertures, and the potential reduction of the number of subapertures while preserving a similar level of performance as the classically dimensionned AO system [\cite{2022JATIS}]. 
Summing the modal residual variances shown in Figure (\ref{fig:OLrec_WFE_vs_SFreq}) provides an estimation of the total reconstruction WaveFront Error (WFE) for the shifted cases.

\begin{figure}[ht!]
\centering
\includegraphics[width=0.5\textwidth]{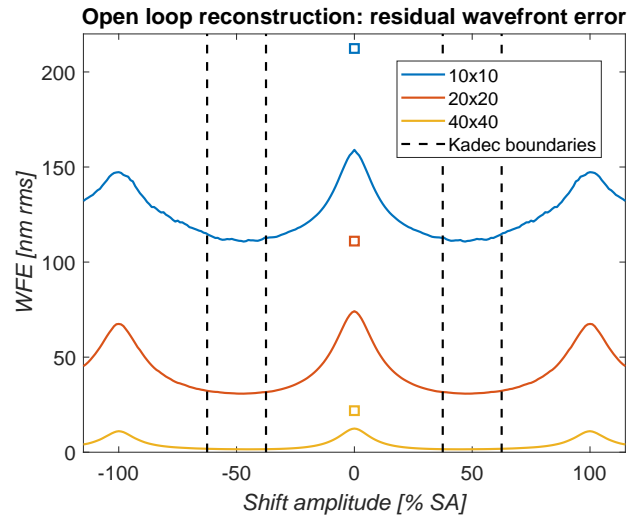} 
\caption{Residual Wavefront Error from open loop reconstruction as a function of shift amplitude for the shifted / super-resolved configurations. In each case, up to four times the native number of modes (4.$\frac{\pi}{4}{n^2_{SA}}$) are reconstructed. The dashed black lines represent the Kadec condition boundaries within which the nonuniform samples can be perfectly reconstructed. The colored squares stand for the reconstruction error from the co-aligned case when $\frac{\pi}{4}{n^2_{SA}}$ modes are reconstructed}
\label{fig:OLrec_WFE_vs_Shift}
\end{figure}

On Fig. (\ref{fig:OLrec_WFE_vs_Shift}), the residual WFE is evaluated as a function of shift amplitude in percentage of a subaperture width. A first observation is the performance gain that can be obtained when reconstructing 4 times more modes than the number of degrees of freedom of a single WFS (colored squares). It can be explained by the \textit{statistical SR} properties. In this case, the MMSE process extrapolates statistically the measurements towards larger spatial frequencies, by assuming the power spectrum described in the Kolmogorov covariance priors. Let us now focus on the \textit{geometric SR} properties. One can observe a periodicity of the achieved WFE with a periodic shift that is a multiple integer of the subaperture width, as expected. However, with any other fractional shift amplitude, the WFE is improved with again an optimal value of half a subaperture that corresponds to a meta uniform sampling generated by the super-resolved system. At this optimal offset of half a subaperture width, it is interesting to note that the 10x10 (resp. 20x20) shifted configuration achieves a WFE that tends towards that from a 20x20 (resp. 40x40) co-aligned system with a number of reconstructed modes equal to the native number of degrees of freedom ($\frac{\pi}{4}{n^2_{SA}}$).   

As already pointed out, the \textit{geometric SR} provides an improvement of the WFE as soon as there is a small offset introduced. So, even non-optimal shifts induce a dramatic improvement of the reconstruction quality. Furthermore, the stability of the WFE around the optimal shift values is remarkable. In summary, Fig. (\ref{fig:OLrec_WFE_vs_Shift}) confirms on one hand that the meta uniform sampling is optimal as demonstrated in Appendix \ref{APP}. On the other hand, it appears unnecessary to accurately tune the system with such optimal offsets since the performance appears weakly sensitive to the offset value. Actually, this observation is consistent with the Kadec condition [\cite{Kad64}], which corresponds to a range of shifts of $+/- 12.5 \%$ of a subaperture width around the half subaperture offset, within which the samples can be perfectly reconstructed. In our simulation, the MMSE reconstructor was not optimized for each shift value, hence there is a slight performance degradation when going away from the meta uniform sampling case. The Kadec range can be interpreted as a tolerance to uncontrolled misalignments, as long as these "misregistrations" are calibrated or known. This feature opens the door to a diversity of geometrical transformation breaking the symmetries of the wavefront sampling topology.
Furthermore, [\cite{Agapito2020}] suggests that SR is more robust to model calibration errors than "classical" AO system of the same dimension.

In the case of a multi WFS tomographic AO system, the super-resolution is "built-in" when it comes to the turbulent layers conjugated in altitude.  Indeed, when projecting the lenslet array grids onto the altitude layers, the off axis WFS grids are shifted proportionally to the  conjugation altitude. It is straight forward that there are some layers altitude where the resulting shift is a multiple integer of the sampling step hence no super-resolution is granted in this case. An optimized geometrical arrangement with possibly a rotation of the lenslet array grids with respect to each other allows the mitigation of this effect.

\begin{figure}[ht!]
\centering
\includegraphics[scale=0.5]{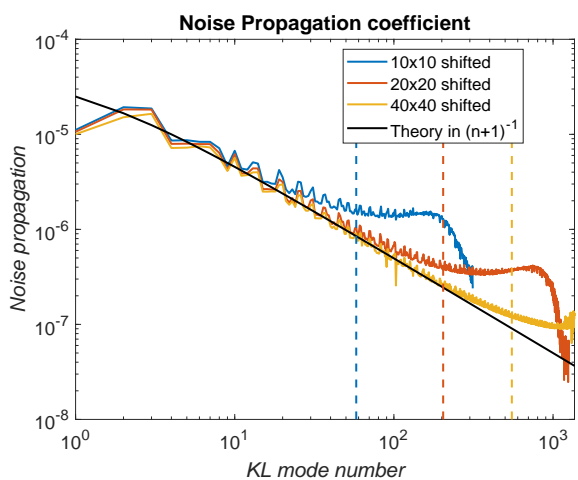} 
\caption{Noise propagation coefficients, computed as $diag \{R R^T \}$, shown in logarithmic scale and scaled by the flux collecting surface of the subaperture.The shifted cases are plotted with solid lines. For the sake of visibility, dashed lines are overlayed at the level of the spatial frequencies where the aliasing starts being prominent for the co-aligned cases. Along the low order modes, the noise propagation is quasi identical for all cases and fit the expected law in $(n+1)^{-1}$ . Then, when exceeding the original Nyquist-Shannon cut-off frequency, the noise propagation coefficient converges towards a constant asymptotic limit.}
\label{fig:noisePropagation}
\end{figure}

Finally, let us evaluate how SR behaves in terms of noise propagation.  Figure (\ref{fig:noisePropagation}) -- which is essentially a numerical confirmation of Fig. (\ref{fig:noisePropSpatFreqs}) -- depicts the noise propagated as a function of the KL mode number for a combination of four i) co-aligned and ii) optimally shifted by half a subaperture width SH-WFS. 
The noise propagation follows the expected $(n+1)^{-1}$ power law with the shifted system (on which SR applies) not exceeding the co-aligned case. We note that for each of the tested cases, the noise propagation within the original cut-off frequency range is almost indistinguishable from that of the co-aligned system whereas past the original Nyquist-Shannon cut-off frequency it rolls off, asymptotically saturating to the  $(n+1)^{0}$ regime. 

\section{Super-resolution WFR with Pyramid WFSs}\label{PWFS}

We argue that the Pyramid WaveFront Sensor (P-WFS) also benefits
from an effect
akin to the SR in the Shack-Hartmann WFS
(SH-WFS). Unlike the latter, the SR principle can be directly applied to a single P-WFS when we allow a more general formalism than Eq. (\ref{COMB}) whereby the samples are obtained from transformations of the initial function. In the case of the P-WFS, each quadrant applies a different transformation, each required for proper signal reconstruction. However, two diagonally opposite quadrants measure the same information for the spatial frequencies larger than the modulation radius [\cite{Guyon2005}]. Thus, the mono-dimensional SR formalism presented in this paper (see \S \ref{PRINC})  can in principle be applied to the corresponding modes for each pair of quadrants whose pixel grids are shifted in one direction by a fraction of a pixel.

Here, the multiple samples are nominally four
diffraction patterns produced at the corresponding re-imaged pupil planes, yet
with the sampling grid pixels shifted with respect to one another in
such a way that their sampling becomes
interlaced by a fractional pixel.  By adjusting the P-WFS apex angles
one such configuration can be readily obtained. The same can in principle be done simply by rotating the pyramid optic with respect to the detector. 
This relaxes the specification and manufacturing constraints, yet in order to obtain
the optimal SR-enabling configuration we may end up re-introducing anew such design constraints we ought to relax in the first place. Clearly, more investigation on the optimality and robustness are required and we invite the community to help in this endeavour. 

To exploit the SR enhancement however, the use of the four diffraction intensity patterns
(i.e. \textit{intensity-maps}) instead
of the customarily \textit{slopes-maps} seems required. 

Is is known
that the 'x' and 'y' slopes-maps contain the complete
information on perfectly aligned systems, i.e.  exactly registered
quadrants and a perfect 4-sided prism with homogeneous and stationary
pupil illumination [\cite{deo18a}]. Yet, additional slopes-maps are
required when these conditions fail. In our opinion the "extension''
proposed in [\cite{deo18a}] is unnecessary provided that it represents a
full-rank linear combination of diffraction intensity patterns. This intermediate step adds nothing to the use of intensity-maps
directly since it neither adds nor removes useful information to that originally contained in the pixels directly. Furthermore it can only be detrimental to the
pipelined pixel processing creating additional idle time before
matching pixels become available during detector read-out. 

Mathematically, out of the eight possible linear combinations (we can take each intensity map either positive or negative and there are four of these), one full-rank slopes-map is
\begin{equation}
  \left[ \begin{array}{c}
           s_x \\ s_y \\ s_{xy} \\ s_t
         \end{array} \right]
     = 
\left[ \begin{array}{cccc}
         1 & -1 & 1 & -1 \\
         1 & 1& -1 & -1 \\
         1 & -1 & -1 & 1 \\
         1 & 1 & 1& -1
       \end{array} \right]
  \left[ \begin{array}{c}
           i_1 \\ i_2 \\ i_3 \\ i_4
         \end{array} \right]   / F_t
\end{equation}
where $F_t$ is a scalar value that represents the total flux captured on all the four re-imaged pupils. We have made a slight change to the last entry $s_t$ which in [\cite{deo18a}] is the total flux $s_t = \left[ 1,  1, 1, 1\right]   \left[ i_1, i_2,
    i_3, i_4\right]^\T$ but leads essentially to the same results for
    the rank of either transformation matrix is 4. 

    As a matter
of fact we go beyond [\cite{deo18a}] in that we do not restrict
ourselves to developing a misalignment  mitigation strategy. Under the
SR framework we claim that
by expressly misaligning the PWFS system we can enhance sensitivity
and therefore gain in the delivered optical performance.

The use of intensity-maps has obvious implications for real-time processing (i.e. twice as large reconstruction
matrix) that needs to be carefully counter-balanced against the
performance gains.

\begin{figure}[htpb]
	\begin{center}
             \includegraphics[width=0.5\textwidth,
             angle=0]{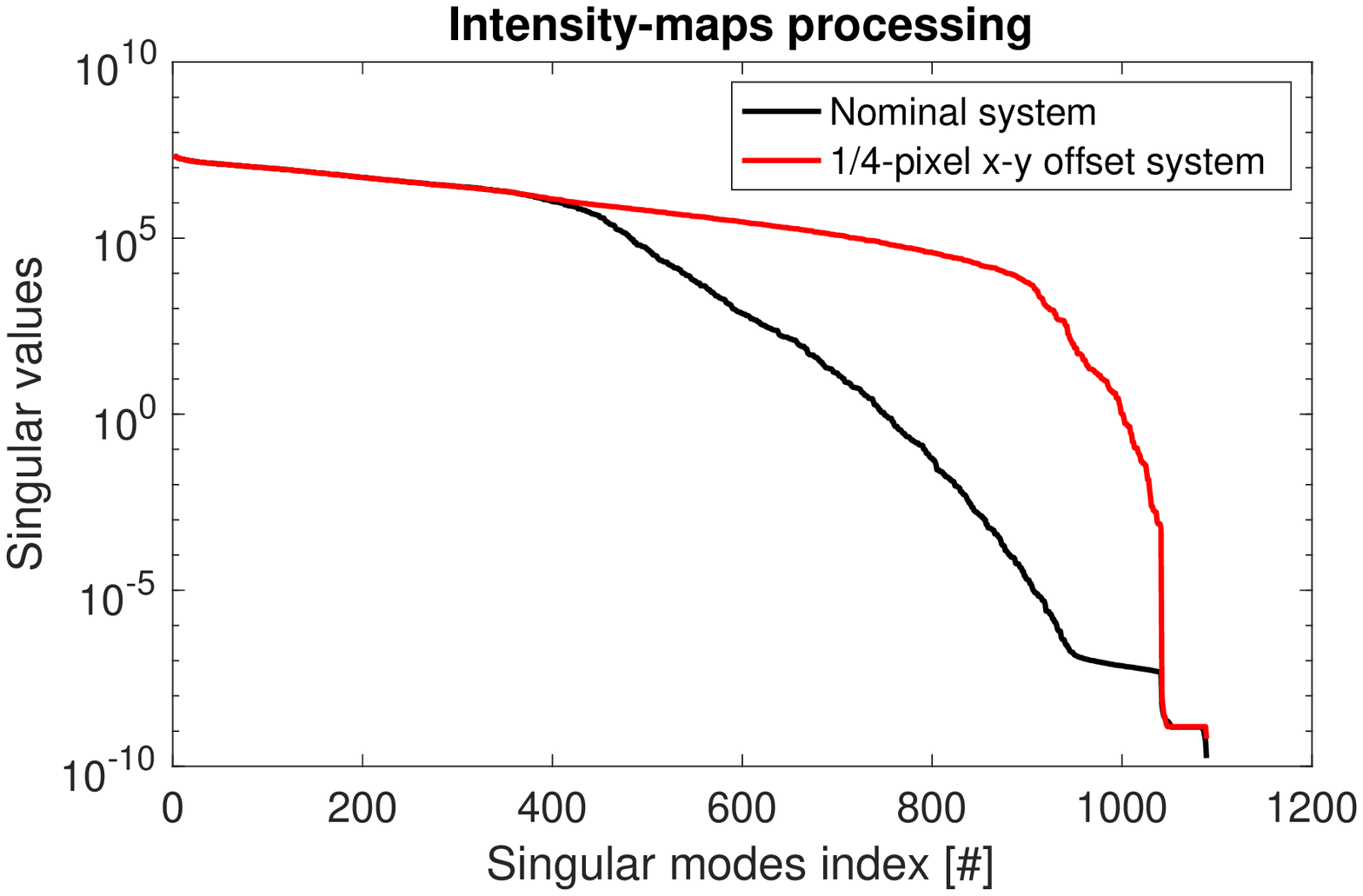}\\\includegraphics[width=0.5\textwidth, angle=0]{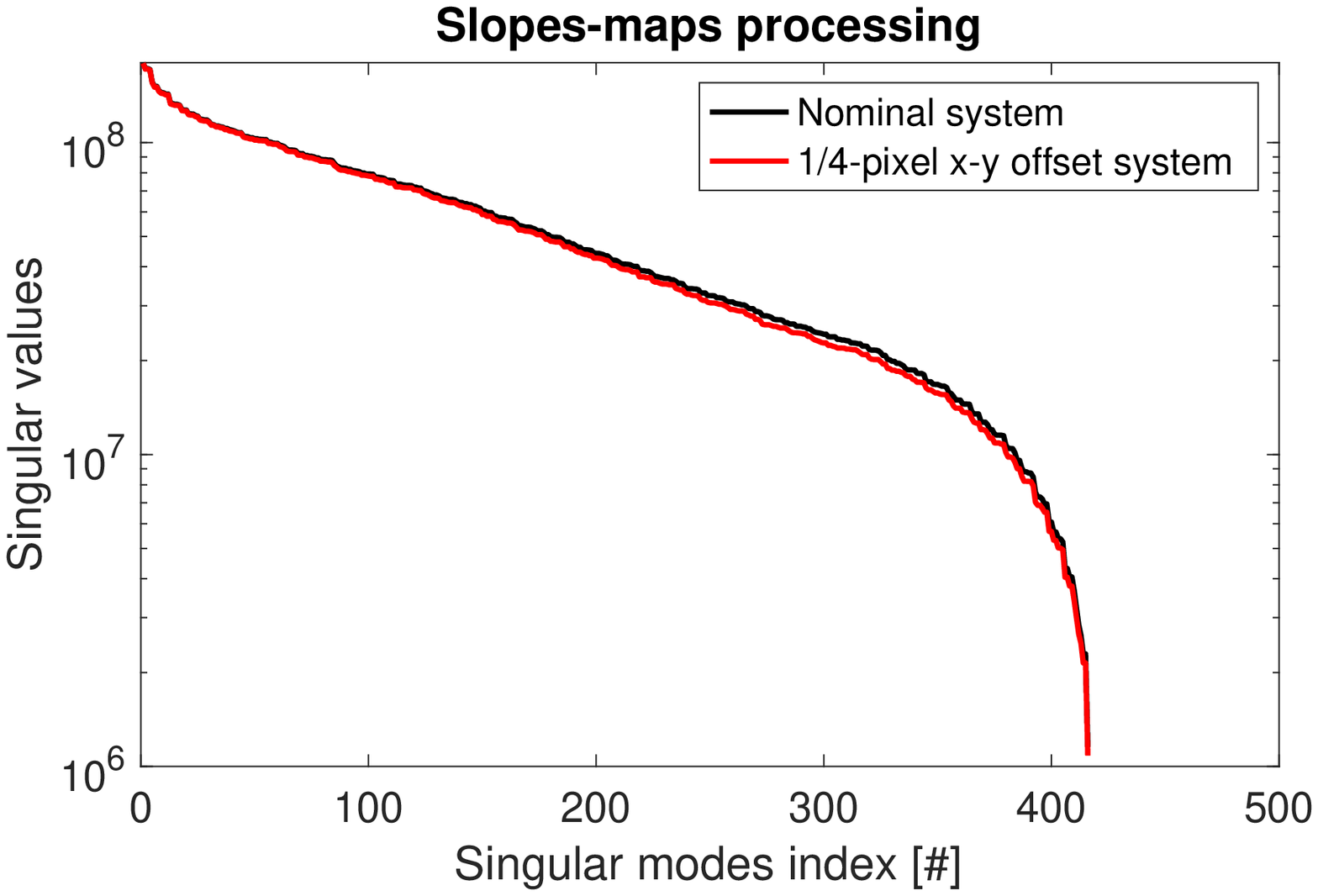}
	\end{center}
	\caption[]
	{\label{fig:svdDecompositionPWFS}
SVD decomposition on a 16x16 16x16 P-WFS system with a 33x33 DM. Top: Intensity-maps processing. Bottom: Slopes-maps processing.  We
clearly see that using the intensity-maps on an offset pixel
sampling system leads to sensitivity gains, whereas the same is not
true were the slopes-maps to be processed.} 
 \end{figure}

To illustrate this behaviour, Fig. (\ref{fig:svdDecompositionPWFS})
shows the SVD decomposition from a 16x16 lenslet based P-WFS system
with a 33x33 regular DM. We clearly see that the
introduction of a relative offset (translation) in the pixels sampling grid
leads to sensitivity gains provided that the intensity-maps are processed. The opposite is obtained when the slopes-maps are processed
instead. The mathematical demonstration and interpretation will be
provided in a forthcoming paper. 

\section{Conclusion and outlook}\label{sec:outlook}
Super-resolution is a computational imaging technique that aims to upscale a signal's spatial resolution from multiple lower resolution samples. SR finds in the wavefront reconstruction -- a process at the very heart of astronomical adaptive optics systems -- a particularly fertile setting. Unlike in imaging systems where the ultimate spatial resolution at the focal plane hits the fundamental diffraction limit,  operating in the wavefront domain in or near the pupil plane meets no limiting bound other than the technological ability to sample the wavefront (spatially and temporally) subject to the inescapable SNR concerns. 

To make SR viable, we have identified several ways to introduce unique phase signatures in the transverse plane to the beam propagation, the simpler being a straight offset. As it stands, SR promises to lift instrument design assumptions, e.g. that $n\times n$ measurement samples are needed to drive a $n \times n$ actuators DM, that the registration between the DM and the WFS sampling grid needs to be finely controlled, that the guide-star asterism be regular and a few others, leading to a major overhaul in how the AO community may approach the designs of future instruments. By decoupling the WFS sampling step from the averaging cell size (subaperture width) we are led to a paradigm shift where the same AO control performance can be achieved with fewer WFS sampling points. This in turn leads to less detector pixels, smaller systems and ultimately a lesser computational load. Another strategy is to optimize the performance while keeping the same number of sensing degrees of freedom, by reconstructing modes beyond the cut-off frequency and mitigating the aliasing. Several tomographic AO systems in operation (GALACSI NFM) or in design phase (MAORY, HARMONI LTAO, MAVIS, KAPA, GMT's GLAO and LTAO, etc...) can benefit from the implementation of SR in one way or the other.

With $N$ regular sampling grids of step $h$, the optimal SR sampling scheme is obtained by introducing fractional offsets that are multiple of $\frac{h}{\sqrt{N}}$ in the bi-dimensional case. This way, one generates a \textit{meta sampler} corresponding to the so called meta uniform sampling scheme. However, the nonuniform sampling case is appealing because it relaxes the absolute alignment constraints. Concretely, the location of the grids is flexible. The optimal reconstruction performance can even be reached within the Kadec condition, that is the departure from the meta uniform sampling by less than a quarter of the Nyquist-Shannon period.
Furthermore, we are yet to establish optimal, or at least practical,  SR sampling geometries for future tomographic AO systems with non square number of WFSs. With 6 WFS (MAORY, HARMONI LTAO) or 8 WFS (MAVIS), the solution is not obvious. Preliminary geometry optimization results can be found in [\cite{Cranney2021}]. Some other ways to break the system symmetries are proposed in Appendix \ref{AppB}.

We have shown that SR does not entail larger noise propagation coefficients and is applicable to multi-Shack-Hartmann WFS systems as well as to single Pyramid WFS. In either case, the relative alignment accuracy is traded-off by the precision with which the registration is known to inform the model-based SR reconstruction. We do not exclude -- and actually very much welcome -- data-driven, machine learning algorithms that can further make this trade-off less restricting. Inspiration may be generated by the very fast developments of these techniques in the field of super-resolution applied to imaging. A review of deep learning approaches for image super-resolution is for example provided in [\cite{wang21}].

\begin{acknowledgements} 
This work benefited from the support of the WOLF \& APPLY project ANR-18-CE31-0018 \& ANR-19-CE31-0011 of the French National Research Agency (ANR), the project  Fortalecimiento del Sistema de Investigación e Innovación de la Universidad de Valparaíso UVA20993 and the project ANID/REDES 190071. It has also been prepared as part of the activities of ORP H2020 Framework Programme of the European
Commission’s (Grant number 101004719). Authors are acknowledging the support by the Action Spécifique Haute Résolution Angulaire (ASHRA) of CNRS/INSU
co-funded by CNES, the support of LABEX FOCUS ANR-11-LABX-0013 and the support of the programme d'investissement avenir ANR-21-ESRE-0008.\\

We would like to thank Christophe Verinaud and Guido Agapito for reading parts of the manuscript and providing pertinent comments. Finally, we wish to acknowledge the fruitful discussions with Pierre-Yves Madec, Michel Tallon, Miska Le Louarn and Jason Spyromilio at the inception of this idea back in 2016. Thank you for your critical review and wise feedback. 
\end{acknowledgements}

\bibliographystyle{aa}
\bibliography{references}
\begin{appendix}
\section{Signal reconstruction and optimal sampling grid positions}\label{APP}

For the sake of completeness, we provide in this appendix  mathematical arguments 
that help us estimate the number of sampling grids required for a reconstruction without aliasing, as a function of their relative spatial offsets. This study can be done with simple mathematical tools, if we consider a specific context that we describe bellow. 

We propose to reconstruct a band-limited function $f:\R\to\R$ sampled by a set of $N$  grids with an interpolation formula of the form
\begin{equation}\label{RECONS}
f(x)=\sum_{\ell=1}^{N}\sum_{n=-\infty}^{+\infty}f(u^\ell_n)s_\ell(x-u^\ell_n),
\end{equation}
where for each $\ell\in\{1,\dots, N\}$, the sequence $u^\ell=\{u^\ell_n\}_{n\in\Z}$ gives the points where the function $f$ is sampled. So, we have to find a suitable set of reconstruction functions  $s_1,\dots,s_N$. The existence of such functions depends on the spectral properties of $f$ and on characteristics of the set of sampling grids, such as  their number, their step and their positions. A restriction lies in the fact that formula \eqref{RECONS} is a convolution. It is a classical hypothesis, but it reduces the set of the possible reconstruction functions  $s_1,\dots,s_N$ and therefore the admissible positions of the grids. However, we will see that formula \eqref{RECONS} leads to $\sinc$ type reconstruction functions, that as such can be implemented at low numerical cost. 

In the context of the PNS framework, we will suppose that all the grids have the same step $h$. However, we are free to set them at any position. So they are of the form
\begin{equation}\label{SSS}
u^\ell_n=nh+u^\ell_0 \quad\forall n\in \Z\quad \ell\in\{1,\dots, N\},
\end{equation}
where $u^\ell_0$ indicates their initial position, which is supposed to be different for each of them.

In the following, we establish the conditions, relating the maximal frequency of $f$ to the characteristics of the sampling sequences, which ensure the existence of the functions $s_1,\dots, s_N$ allowing to reconstruct $f$ with the formula \eqref{RECONS}. 
For a fixed maximal frequency, there are many possible options for the positions of the sampling grids, provided that we own a sufficient quantity of them. However, we are interested in finding the positions which need the smallest number of sampling grids as possible.
As we will see, there is no surprise.  With a set of uniform sampling grids, the meta uniform sampling (Eq. \eqref{FRAC}) is optimal: it allows reaching a given maximal frequency with the lowest number of sampling grids. 

\subsection{Reconstruction conditions}

To find the reconstruction functions $s_1,\dots,s_N$, we will solve Eq. \eqref{RECONS} in the Fourier space. In this context, we assume the hypothesis that all involved functions admit a Fourier transform.
So, we are looking for conditions that ensure the existence of some functions $s_1,\dots,s_N$ such that 
\[
\hat{f}=\hat{g}\quad\text{where}\quad g(x):=\sum_{\ell=1}^{N} \sum_{n=-\infty}^{+\infty}f(u^\ell_n)s_\ell(x-u^\ell_n)\quad\forall x\in\R.
\]  
For the sampling grids that we consider, the Fourier transform of $g$ is given by
\begin{align}
\hat{g}(k)&=C_0(k)\hat{f}(k)+\sum_{n\neq0,n=-\infty}^{+\infty}C_n(k)\hat{f}\left(k-\nu_h n\right)\label{GC},
\end{align}
where
\begin{equation}\label{CN}
C_n(k):=\frac{1}{h}\sum_{\ell=1}^{N}\hat{s}_\ell(k)e^{-2i\pi \nu_hnu^\ell_0}\quad\text{and}\quad\nu_h:=\frac{1}{h}.
\end{equation}
This computation is not direct, but classical and can be found for instance in [\cite{Kohlenberg1953}].

In what follows,  we will suppose  that $f$ is band-limited, hence $\hat{f}(k)$ vanishes outside some symmetric interval $K$. That is, there exists a  frequency $k_{max}>0$  such that   $\hat{f}(k)=0$ for all $k\notin K$, where $K:=(-k_{max},k_{max})$. In most practical situations $\hat{f}$ does not necessarily have a bounded support. But, $\hat{f}(k)$ may decrease rapidly as $|k|$ goes to infinity, as does for instance  the power spectrum density of turbulent wavefronts following a Kolmogorov power law.

The translated copies $\hat{f}(k-\nu_hn)$ of $\hat{f}$, that appear in Eq. \eqref{GC}, may have some non zero values in the set $K$ overlapping $\hat{f}$ in this set, and hence producing aliasing. This happens if for some $n\neq 0$ a set $K_n:=(-k_{max}+\nu_hn,k_{max}+\nu_hn)$ satisfies $K\cap K_n\neq\emptyset$. It is the case if and only if $|n|< {n_0}$, where

\begin{equation}\label{N0}
n_0:=\left\lceil \frac{2k_{max}}{\nu_h}\right\rceil=\left\lceil \frac{k_{max}}{\nu}\right\rceil,
\end{equation}
and $\lceil \cdot \rceil$ is the ceiling  function and $\nu=1/2h$ is the native Nyquist frequency.  
This can be reformulated in the following way: if for some integer $n_0$ the frequency $k_{max}$ satisfies 
$(n_0-1)\nu<k_{max}\leq n_0\nu$, then $K\cap K_n\neq\emptyset$, if and only if $|n|< {n_0}$.  

\begin{figure}[ht!]
\centering
\includegraphics[scale=0.20]{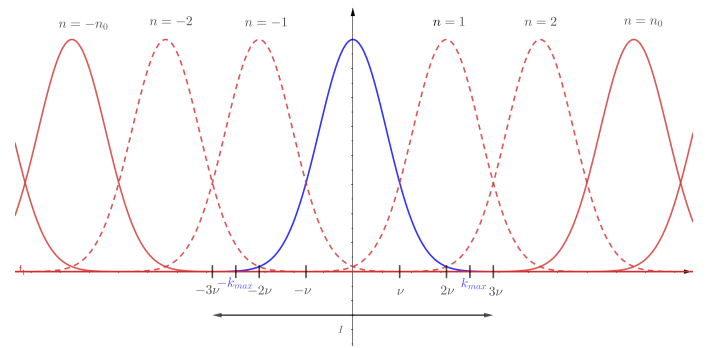} 
\caption{The Fourier transform $\hat{f}$ (blue) and its translated copies (red). Here, $2\nu<k_{max}\leq 3\nu$ and therefore $n_0=3$. The translated copies $\hat{f}(k-\nu_hn)$ overlap $\hat{f}$ for $n\in\{-2,-1,1,2\}$ (dashed curves). On the interval $I=[-3\nu,3\nu]$, there is no contribution of $\hat{f}(k-\nu_hn)$ for any $|n|\geq {n_0}$.}
\label{FIGTF}
\end{figure}

It follows, that in $K$, but also in the bigger set 

\[
I:=[-n_0\nu,n_0\nu]\supset K,
\]
the formula  \eqref{GC} writes
 \begin{equation}\label{GN0}
\hat{g}(k)=C_0(k)\hat{f}(k)+\sum_{n\neq0,n=-{n_0+1}}^{{n_0-1}}C_n(k)\hat{f}\left(k-\nu_h n\right)\quad\forall k\in I,
\end{equation}
since $\hat{f}(k-\nu_hn)=0$ for every $k\in I$ if $|n|\geq{n_0}$, see Fig. \eqref{FIGTF}. 

To ensure that $\hat{g}=\hat{f}$ in $I$, and therefore in $K$, we can impose for every $k\in I$ that $C_0(k)=1$
and $C_{n}(k)=0$ for all $n\in\{-{n_0+1},\dots,{n_0-1}\}\setminus\{0\}$, that is,
 \begin{align}
&\sum_{\ell=1}^{N}\hat{s}_\ell(k)=h\qquad\forall k\in I,\label{CONDIN}\\
&\sum_{\ell=1}^{N}\hat{s}_\ell(k)e^{2i\pi\nu_hnu^\ell_0}=0 \quad\forall\,n\in\{-{n_0+1},\dots,{n_0-1}\}\setminus\{0\}\quad\forall k\in I.\label{CONDHOM}
\end{align}
Now, for $k\notin I$, we have $\hat{f}(k)=0$ and therefore we want $\hat{g}(k)=0$. 
To cancel the contributions of all the translated copies of $\hat{f}$ outside of $K$, we can simply set
\begin{equation}\label{CONDOUT}
\hat{s}_1(k)=\dots=\hat{s}_N(k)=0\qquad \forall k\notin I.
\end{equation}

\noindent{\bf Remark:} The Eq. \eqref{CONDIN} and Eq. \eqref{CONDHOM} are sufficient conditions to obtain $\hat{g}=\hat{f}$ in the set $I$ and to  reconstruct $f$ using formula \eqref{RECONS}. 
However, it may be possible to propose weaker but more elaborated conditions.
For the sake of simplicity,  we will restrict the study to Eq. \eqref{CONDIN} and Eq. \eqref{CONDHOM}.  

\medskip
Let us sum up our results and introduce  more suitable notations for the forthcoming study of the system given by Eq. \eqref{CONDIN} and Eq. \eqref{CONDHOM}. We consider a band-limited function $f$, whose Fourier transform vanishes outside of  $(-k_{max},k_{max})$, and $N$ sampling  sequences of step $h$ and initial positions $u_0^1,\dots u_0^N$ (see Eq. \eqref{SSS}). If we  introduce the quantities
\[
\beta_\ell:=e^{2i\pi\nu_hu^\ell_0}=e^{2i\pi u^\ell_0/h}\quad\forall\, \ell\in\{1,\dots, N\},
\]
then the system defined by Eq. \eqref{CONDIN} and Eq. \eqref{CONDHOM} is of the form  $AX=Y$, where $X$ is a vector of size $N$,  $A$  is the $(2{n_0-1})\times N$ matrix defined by
\begin{equation}
A:=\begin{pmatrix}\label{SYSTEM}
\beta_1^{-{n_0+1}} & \beta_2^{-{n_0+1}} & \cdots & \beta_N^{-{n_0+1}} \\

\vdots  & \vdots  & \ddots & \vdots  \\
1 & 1 & \cdots & 1\\ 
\vdots  & \vdots  & \ddots & \vdots  \\
\beta_1^{{n_0-1}} & \beta_2^{{n_0-1}} & \cdots & \beta_N^{{n_0-1}} 
\end{pmatrix}
\quad\text{and}\quad
Y:=\begin{pmatrix}
0 \\
\vdots  \\
h\\ 
\vdots \\
0 
\end{pmatrix}.
\end{equation}
If $X=(x_1,\dots,x_N)$ is a solution of the system \ref{SYSTEM}, and $\hat{s}_1\dots, \hat{s}_N$ are defined for any $\ell\in\{1,\dots, N\}$ by
\begin{equation}\label{FORMSC}
\hat{s}_\ell(k)=x_\ell \quad\forall k\in I\quad\text{and}\quad \hat{s}_\ell(k)=0  \quad\forall\, k\notin I,
\end{equation}
then  $\hat{g}=\hat{f}$. It follows that Eq. \eqref{RECONS} holds, with the reconstruction functions: 
\begin{equation}\label{FREC}
s_\ell(x)=\frac{x_\ell n_0}{h}\sinc\left(\frac{\pi n_0x}{h}\right)\quad\forall\,\ell\in\{1,\dots, N\}.
\end{equation}
Let us give a simple example
and re-prove at the same time the Nyquist-Shannon Theorem.\\

\noindent {\bf Example:} By the Nyquist-Shannon Theorem, we know that, if $k_{max}\leq 1/2h = \nu$, a  sampling sequence of step $h$  allows to perfectly reconstruct $f$. We can re-prove that result. Indeed, if $k_{max}\leq 1/2h$, then formula \eqref{N0} gives ${n_0}=1$ and the system \eqref{SYSTEM} to solve writes
\[
x_1+\dots+x_N=h.
\]
If $N\geq 2$, we can produce an infinite number of solutions of this equation, and we have as many possible choices for $\hat{s}_1\dots, \hat{s}_N$. But a single sampling grid ($N=1$),  is enough to ensure the existence of a solution, which  is $x_1=h$. So, let   
\begin{equation*}
\hat{s}_1(k)=h \quad\forall\, k\in I\quad\text{and}\quad \hat{s}_1(k)=0 \quad\forall\, k\notin I,
\end{equation*}
where $I=[-1/2h,1/2h]$.
Applying Eq. \eqref{FREC} and using Eq. \eqref{RECONS}, we obtain 
\[
s_1(x)=\sinc(\pi x/h)\quad\text{and} \]
\[
f(x)=\sum_{n=-\infty}^{+\infty}f(nh+u_0^1)\sinc\left(\frac{\pi}{h}(x-nh-u_0^1)\right),
\]
as expected.\\

\noindent{\bf Remark:} In what follows we will study the conditions on $N$ and $u_0^1,\dots,u_0^N$ that ensure the existence of a solution to the system \eqref{SYSTEM}. As explained, if such a system has a solution then we can  reconstruct $f$ using formula \eqref{RECONS}. If the system does not have solution, then there is no functions $\hat{s_1},\dots,\hat{s_1},$ that satisfies Eq. \eqref{CONDIN} and Eq. \eqref{CONDHOM}. This does not necessarily mean that $f$ cannot be reconstructed with formula \eqref{RECONS}, since  Eq. \eqref{CONDIN} and Eq. \eqref{CONDHOM} are only sufficient conditions to obtain $\hat{g}=\hat{f}$.

\subsection{Optimality of the uniform sampling}
The system \eqref{SYSTEM} has $2{n_0-1}$ equations with $N$ unknowns. The number of equations is imposed by the highest frequency $k_{max}$ of the signal $f$ and by the step $h$ of the sampling grids, through the formula \eqref{N0}. The number of unknowns is equal to the number $N$ of sampling grids that we own. The coefficients of the system depend on $h$ and can be tuned by our choice of the initial positions $u_0^1,\dots,u_0^N$ of the sampling grids. Here, we are interested in the initial positions which ensure the existence of a solution 
to Eq. \eqref{SYSTEM}, with the smallest number of sampling grids $N$ as possible. 

\subsubsection{Minimal number of sampling grids}

If $N=2{n_0-1}$, then $A$ is a square matrix with a Vandermonde determinant. This determinant vanishes if and only if $\beta_\ell=\beta_{l'}$ for some $\ell\neq \ell'$. This occurs if and only if $u_0^{\ell}=u_0^{\ell'}+qh$ for some $q\in\Z$, that is, if the two grids $\ell$ and $\ell'$ have the same points (see Eq. \eqref{SSS}). Therefore, if the sampling grids are all different, then the matrix $A$ is invertible. In such a case, the system \eqref{SYSTEM} has  a solution, which is unique for each possible position of $u_0^1,\dots,u_0^N$. We can then reconstruct our signal choosing the functions $\hat{s}_1,\dots,\hat{s}_N$ according to Eq. \eqref{FORMSC}. We conclude that with $N=2n_0-1$ grids, we can reconstruct a signal such that  $(n_0-1)\nu <k_{max}\leq n_0\nu$, even if the initial positions of the grids are chosen randomly.  

However, for well chosen positions of $u_0^1,\dots,u_0^N$, we can reduce the number of independent equations of the system and obtain a solution even if $N<2{n_0-1}$. Indeed,
let us suppose that we have $N={n_0}$ sampling grids of the form 
\begin{equation}\label{FRACG}
u^\ell_n=nh+u^1_0+(\ell-1)\frac{h}{N}\quad\forall \ell\in\{1,\dots,N\}.
\end{equation}
Then, for every $\ell\in\{1,\dots,N\}$ and $j\in\{-{n_0+1},\dots,-1\}$, we have
\[
\beta_\ell^{j+n_0}=e^{2i\pi(j+N)u^\ell_0/h}=\beta_\ell^{j}e^{2i\pi Nu_0^1/h}e^{2i\pi(l-1)}=\beta_1^{N}\beta_\ell^{j}.
\]
This implies that to multiply the ${n_0-1}$ first lines of $AX=Y$ by $\beta_1^{N}$ gives a system where each of the ${n_0-1}$ first equations is equal to one of the ${n_0-1}$ last equations. Therefore, if $X=(x_1,\dots,x_N)$ is a solution of 
\begin{equation}\label{SRED}
\begin{pmatrix}
1 & 1 & \cdots & 1\\ 
\beta_1 & \beta_2 & \cdots & \beta_N\\ 
\vdots  & \vdots  & \ddots & \vdots  \\
\beta_1^{{n_0-1}} & \beta_2^{{n_0-1}} & \cdots & \beta_N^{{n_0-1}} 
\end{pmatrix}
\begin{pmatrix}
x_1\\ 
x_2\\ 
\vdots \\
x_N 
\end{pmatrix}
=\begin{pmatrix}
h\\ 
0\\ 
\vdots \\
0 
\end{pmatrix},
\end{equation}
then $X$ is also a solution of $AX=Y$. The square system \eqref{SRED} has a Vandermonde determinant. Since the points of the grids  \eqref{FRACG} are all different, the system \eqref{SRED} is invertible. We can verify that its unique solution is given by $x_1=x_2=\dots=x_N=h/N$. 
So, we re-prove  that with $N=n_0$ grids chosen according to Eq. \eqref{FRACG}, we can reconstruct a signal such that  $(n_0-1)\nu <k_{max}\leq n_0\nu$.  

\medskip
Now, we wonder if there is some smarter way to select the positions of the sampling grids, which allows to reduce even more their number. The answer is negative. Indeed, 
if $N\leq {n_0-1}$, we can extract from Eq. \eqref{SYSTEM} the subsystem
\begin{equation}\label{SYSHOM}
\begin{pmatrix}
\beta_1& \beta_2 & \cdots & \beta_N \\
\vdots  & \vdots  & \ddots & \vdots  \\
\beta_1^{N} & \beta_2^{N} & \cdots & \beta_N^{N} 
\end{pmatrix}
\begin{pmatrix}
x_1\\ 
\vdots \\
x_N 
\end{pmatrix}
=\begin{pmatrix}
0\\ 
\vdots \\
0 
\end{pmatrix},
\end{equation}
which must be satisfied by any solution  of Eq. \eqref{SYSTEM}. As Eq. \eqref{SYSHOM} is an invertible homogeneous system, its unique solution is $X=(0,0,\dots,0)$, which is not a solution of Eq. \eqref{SYSTEM}.
So Eq. \eqref{SYSTEM} has no solution if $N\leq n_0-1$. 

\subsubsection{Unicity}

We have shown that $N={n_0}$ is the minimal number of sampling grids for the existence of a solution to the system \eqref{SYSTEM}.  In particular, if we choose our  grids according to Eq. \eqref{FRACG} it has  a solution.  Now, we prove that for any other choice of the position of $n_0$ grids, the system \eqref{SYSTEM} has no solution.\\

To that aim, we look for a necessary condition for the rank of $A$ to be equal to the rank of $A|Y$, where $A|Y$ is the matrix $A$ with an extra row equal to $Y$. First, we have $\rank(A)= N$, since on one hand the number of rows $N=n_0$ of $A$ is smaller than its number of lines $2{n_0-1}$, and on the other hand $A$ contains a $N\times N$ Vandermonde invertible matrix. So, it is enough to look for a necessary condition for  $\rank(A|Y)= N$. 

As $N={n_0}$,   $A|Y$ is a $(2N-1)\times(N+1)$ matrix, from which  we can extract a collection of $N-1$ matrices $B_1,\dots,B_{N-1}$ of size $(N+1)\times(N+1)$ and defined by:

\[
B_i:=\begin{pmatrix}
\beta_1^{-i} & \beta_2^{-i} & \cdots & \beta_N^{-i} & 0\\
\vdots  & \vdots  & \ddots & \vdots & \vdots  \\
1 & 1 & \cdots & 1& h\\

\vdots  & \vdots  & \ddots & \vdots & \vdots  \\
 
\beta_1^{N-i} & \beta_2^{N-i} & \cdots & \beta_N^{N-i} & 0
\end{pmatrix}
\qquad\forall\, i\in\{1,\dots, N-1\}.
\]
We can have $\rank(A|Y)= N$, only if the $\det(B_i)=0$ for all $i\in\{1,\dots, N-1\}$. A Laplace expansion along the last row gives   
\[
\det(B_i)=(-1)^{N+i+2}h\det(C_i)\prod_{\ell=1}^{N}\beta^{-i}_\ell \quad \text{where}
\]
 
\[
C_i:=
\begin{pmatrix}
1 & 1 & \cdots & 1\\
\vdots  & \vdots  & \ddots & \vdots   \\

\beta_1^{i-1} & \beta_2^{i-1} & \cdots & \beta_N^{i-1} \\
\beta_1^{i+1} & \beta_2^{i+1} & \cdots & \beta_N^{i+1} \\

\vdots  & \vdots  & \ddots & \vdots  \\

\beta_1^{N} & \beta_2^{N} & \cdots & \beta_N^{N} 
\end{pmatrix}\quad \forall\, i\in\{1,\dots, N-1\}.
\]
Let $P(X)=a_NX^N+\dots +a_iX^i+\dots+a_0$ be the polynomial defined by 
\[
P(X):=
\begin{vmatrix}
1 & 1 & \cdots & 1 & 1\\
\vdots  & \vdots  & \ddots & \vdots&  \vdots \\
\beta_1^i& \beta_2^i & \cdots & \beta_N^i & X^i\\

\vdots  & \vdots  & \ddots & \vdots &  \vdots\\
 \beta_1^{N} & \beta_2^{N} & \cdots & \beta_N^{N} & X^N 
\end{vmatrix}.
\]
Then, a Laplace expansion along the last row gives $a_i=(-1)^{i+N+1}\det(C_i)$ for all $i\in\{1,\dots, N-1\}$. Therefore, we have $\det(B_i)=0$ for all $i\in\{1,\dots, N-1\}$ if and only if
\[
P(X)=a_NX^N+a_0.
\]
The coefficients $a_N$ and $a_0$ are obtained by Laplace expansion, and if we denote $V_{N}(\beta_1,\dots,\beta_N)$  the Vandermonde determinant of size $N$, then 

\[
P(X)=V_{N}(\beta_1,\dots,\beta_N)\left(X^N+(-1)^{N+2}\prod_{\ell=1}^{N}\beta_\ell\right). 
\]
On the other hand, we have $P(\beta_\ell)=0$ for all $\ell\in\{1,\dots, N\}$, which gives
\[
\beta_1^N=\beta_2^N=\dots=\beta_N^N=(-1)^{N+1}\prod_{\ell=1}^{N}\beta_\ell.
\]
It follows that $(\beta_\ell/\beta_1)^N=1$ for all $\ell\in\{1,\dots, N\}$. In other words, each  $\beta_\ell/\beta_1$ is a  $N^{th}$ root of the unity. Since all the $\beta_\ell$ are supposed different, we have   
\[
\beta_\ell=\beta_1e^{2i\pi(\ell-1)/N}=e^{2i\pi(u_0^1/h+(\ell-1)/N)}\quad\forall\, \ell\in\{1,\dots,N\}.
\]
Recalling that $\beta_\ell=e^{2i\pi u_0^\ell/h}$, we obtain
\[
u_0^\ell=u_0^1+(\ell-1)\frac{h}{N}\mod h \quad\forall\, \ell\in\{1,\dots,N\}.
\]
We conclude that a necessary condition for the system \eqref{SYSTEM} to have a solution if $N={n_0}$ is that 
the sampling grids satisfy Eq. \eqref{FRACG}.

\subsubsection{Conclusion}
To sum up, with $N= 2{n_0-1}=2\left\lceil{k_{max}}/{\nu}\right\rceil-1$ sampling grids of step $h$, we can always reconstruct 
a signal $f$ with maximal frequency  $k_{max}$. The sampling grids positions can be chosen randomly, provided that the grids are all different. However, the same performance can be reached with only ${N=n_0}$ sampling grids, if their union is a uniform sampling grid of step $h/N$, that is, if they satisfy Eq. \eqref{FRACG} modulo a translation of size $h$. On the other hand, if we own less than ${n_0}$ sampling grids, or ${n_0}$ sampling grids that do not satisfy Eq. \eqref{FRACG}, then there is no set of reconstruction functions $s_1,\dots,s_N$ which satisfy Eq. \eqref{CONDIN} and Eq. \eqref{CONDHOM}. Since these conditions could be too strong, we cannot conclude from this study that the signal cannot be reconstructed with more elaborated approaches and reconstruction functions.  This study suggests nevertheless that the meta uniform sampling Eq. \eqref{FRACG} is optimal, because it allows to reconstruct the signal without aliasing with the smallest number of sampling grids as possible. 
It is therefore the safest  way to make use of a limited number of sampling grids.

\section{Nonuniform sampling schemes}\label{AppB}

\begin{figure*}[h]
	\begin{center}
            \includegraphics[width=0.35\textwidth,angle=0]{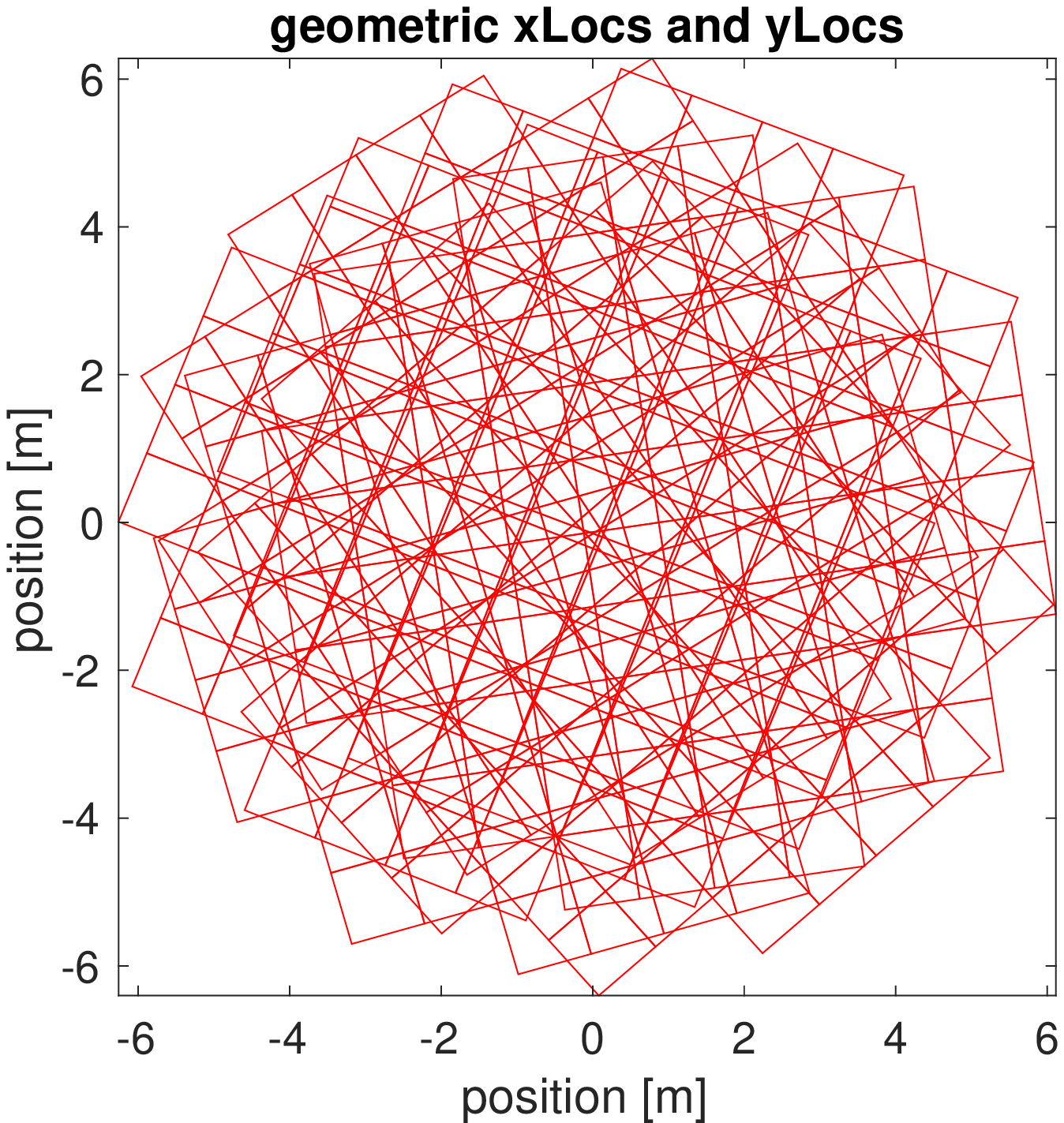}\includegraphics[width=0.35\textwidth,
             angle=0]{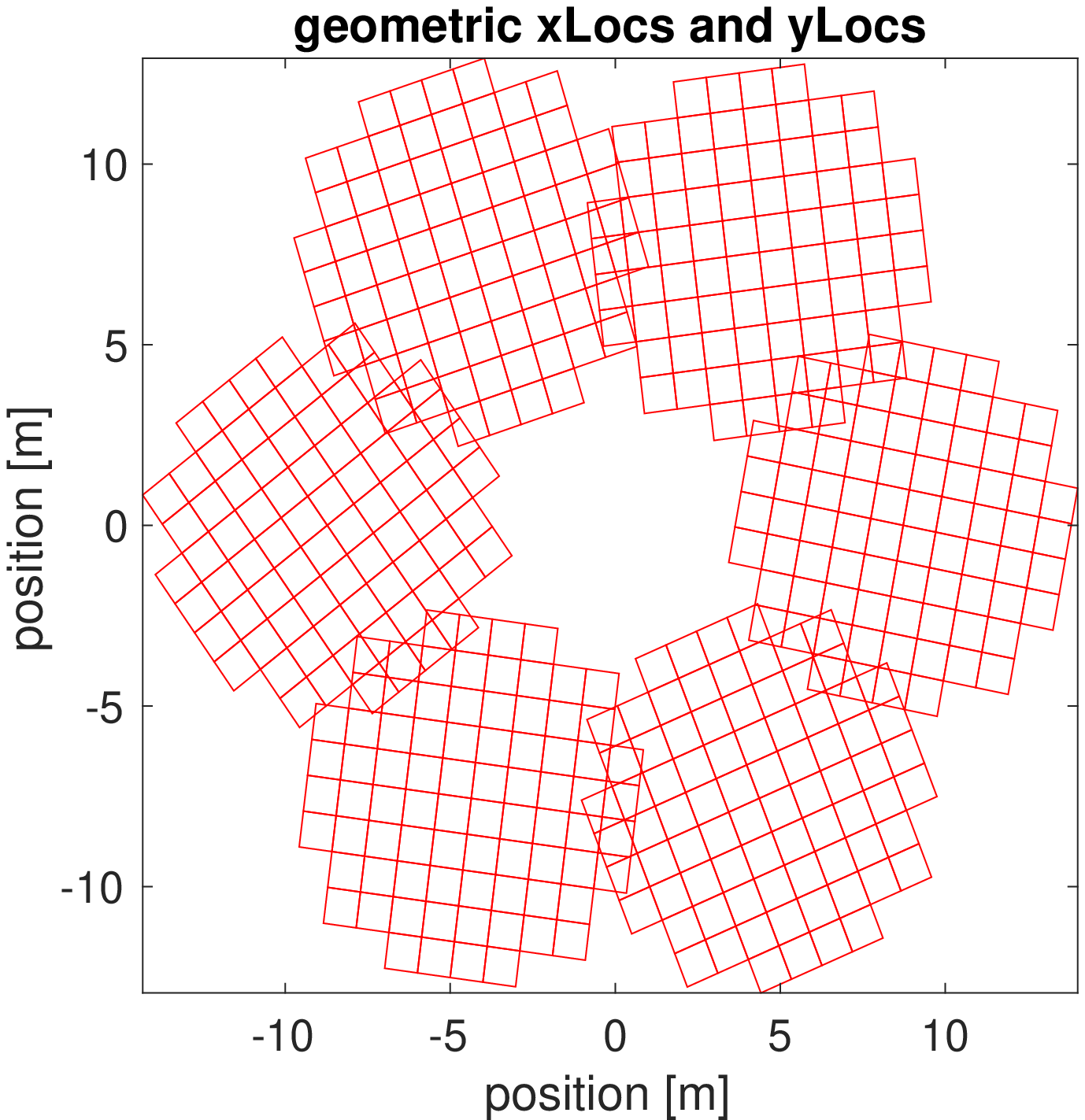}\includegraphics[width=0.35\textwidth,
             angle=0]{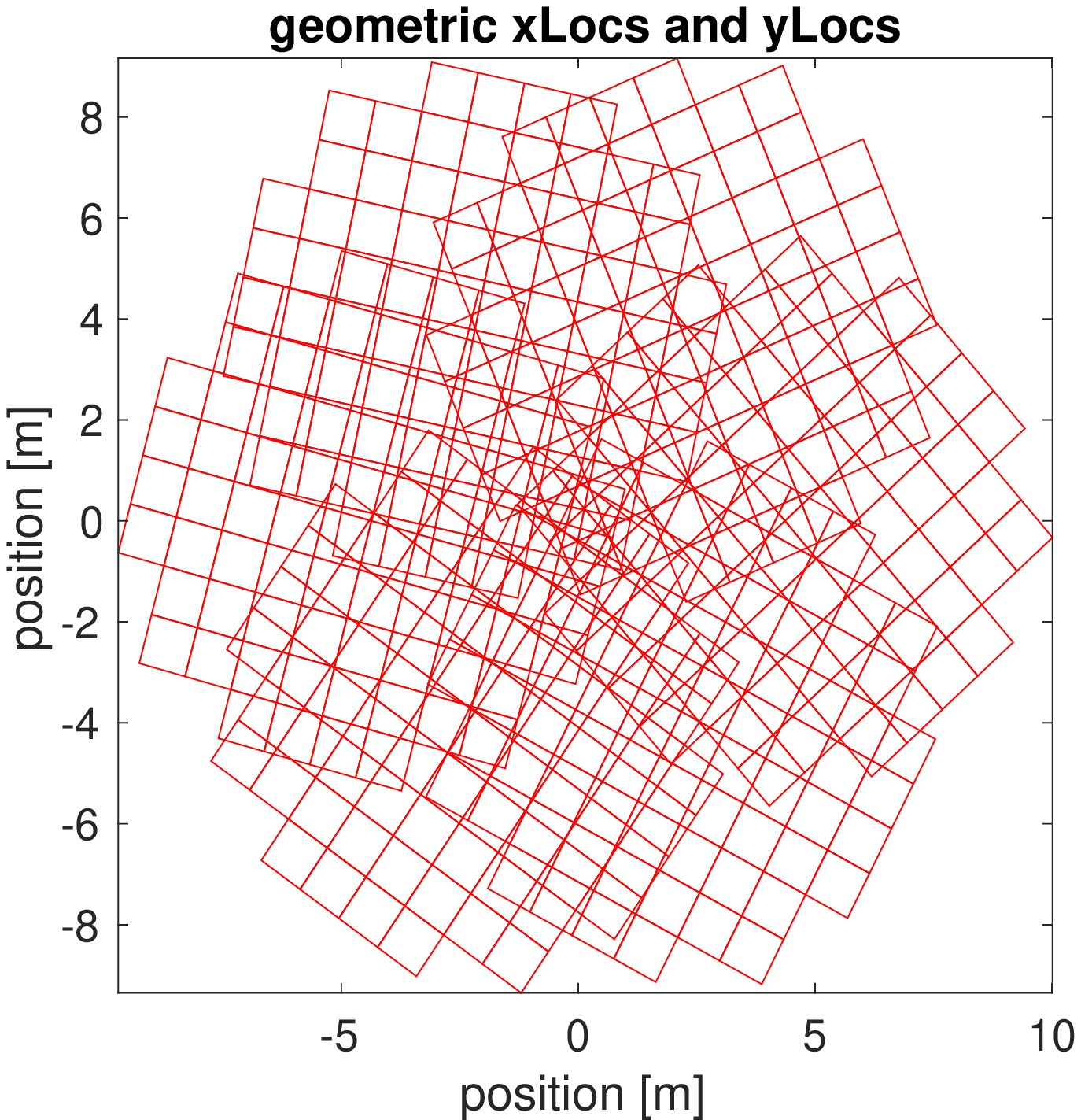}
	\end{center}
	\caption[]
	{\label{fig:hexagonalMeshRandRotGrids1}Six differentially rotated WFSs configuration shown at three different arbitrary conjugation altitudes  creating growing separations (highly exaggerated for illustration only).
 }
\end{figure*}

\begin{figure*}[h]
	\begin{center}
             \includegraphics[width=0.35\textwidth,
             angle=0]{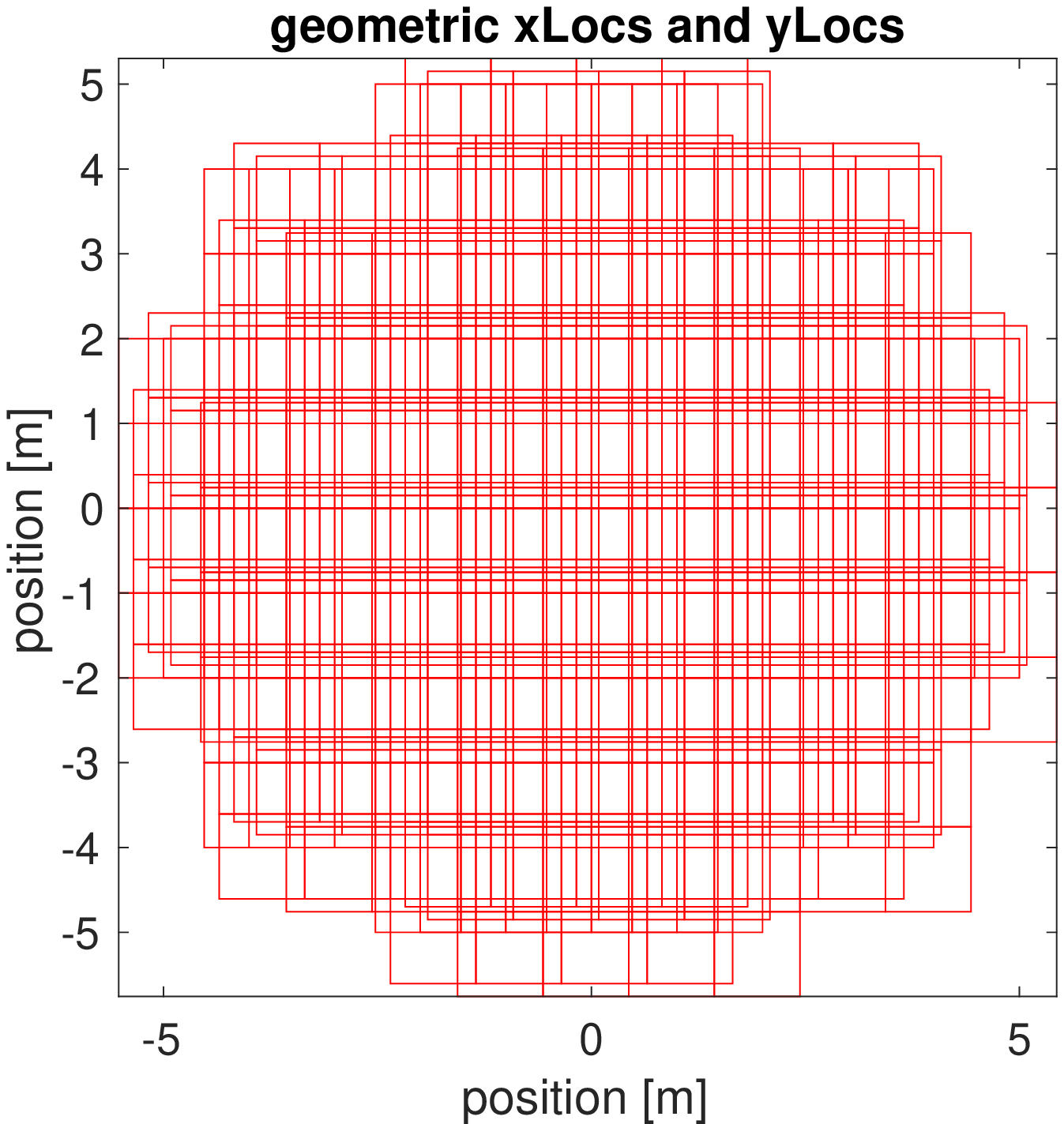}\includegraphics[width=0.35\textwidth,
             angle=0]{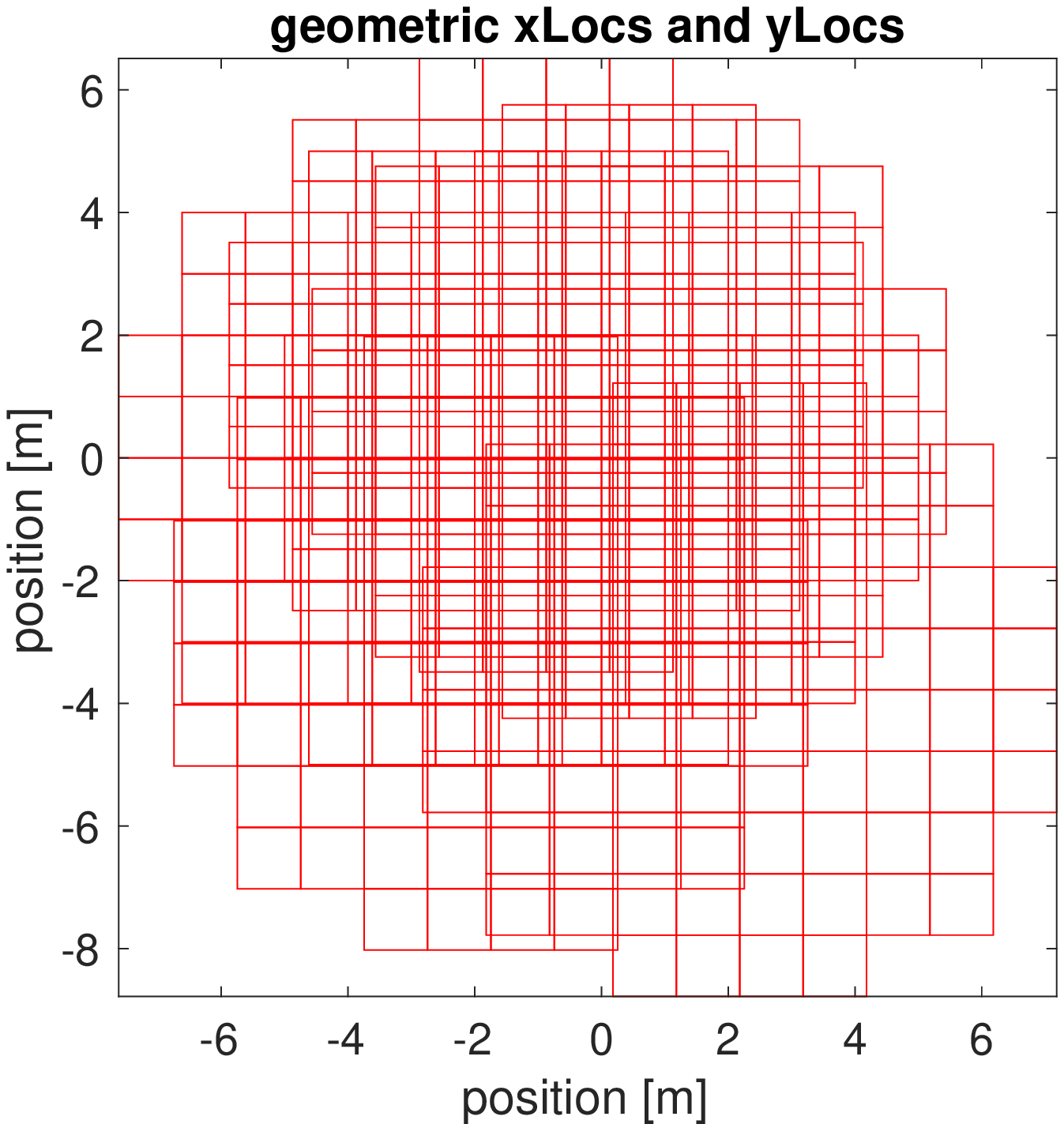}\includegraphics[width=0.35\textwidth,
             angle=0]{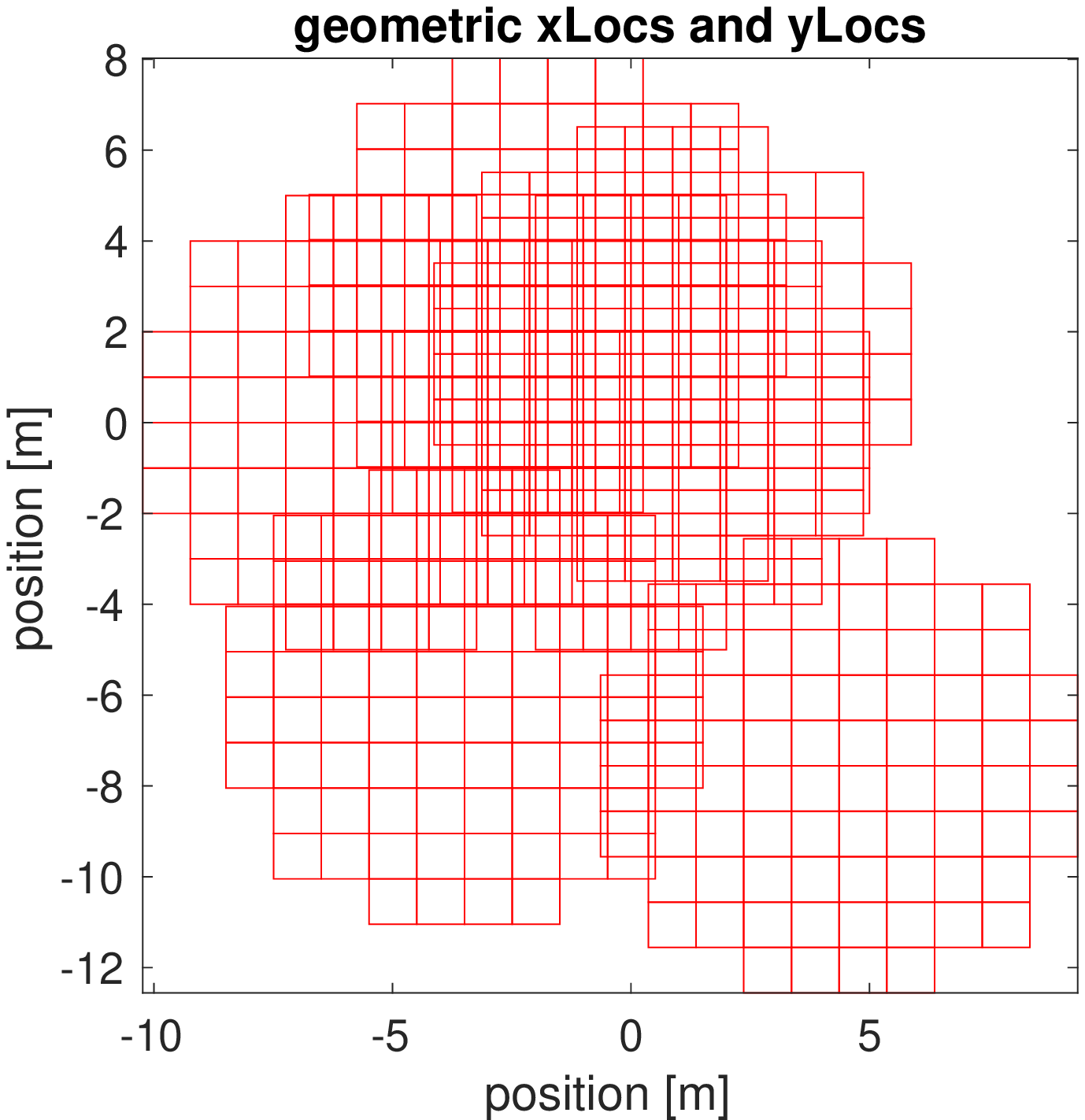}
	\end{center}
	\caption[]
	{\label{fig:hexagonalMeshSpiralGrids1}Six regular WFSs configuration with a spiral shaped asterism shown at three different conjugation altitudes creating growing separations.
 }
\end{figure*}

\begin{figure*}[h]
	\begin{center}
             \includegraphics[width=0.35\textwidth,
             angle=0]{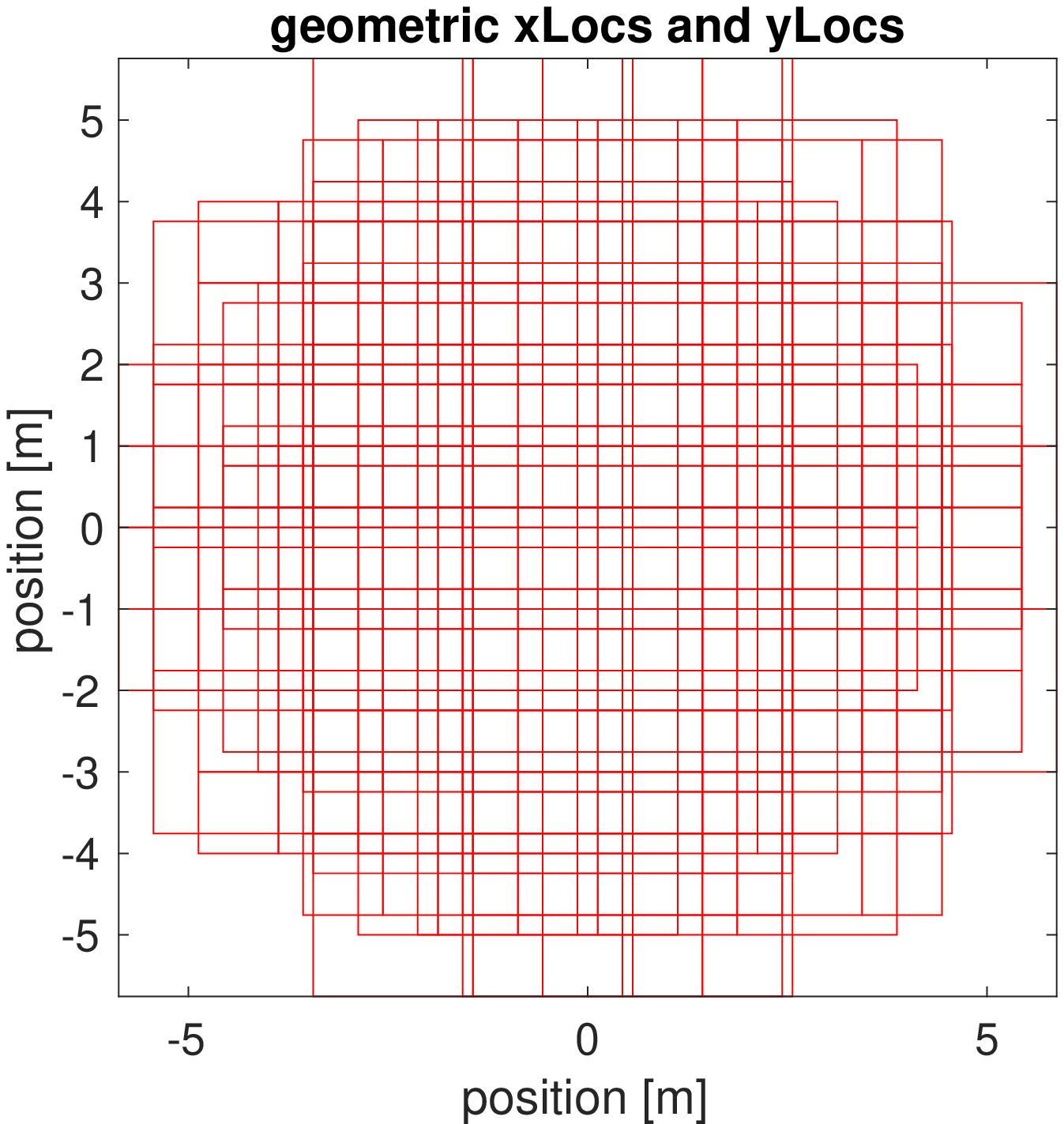}\includegraphics[width=0.35\textwidth,
             angle=0]{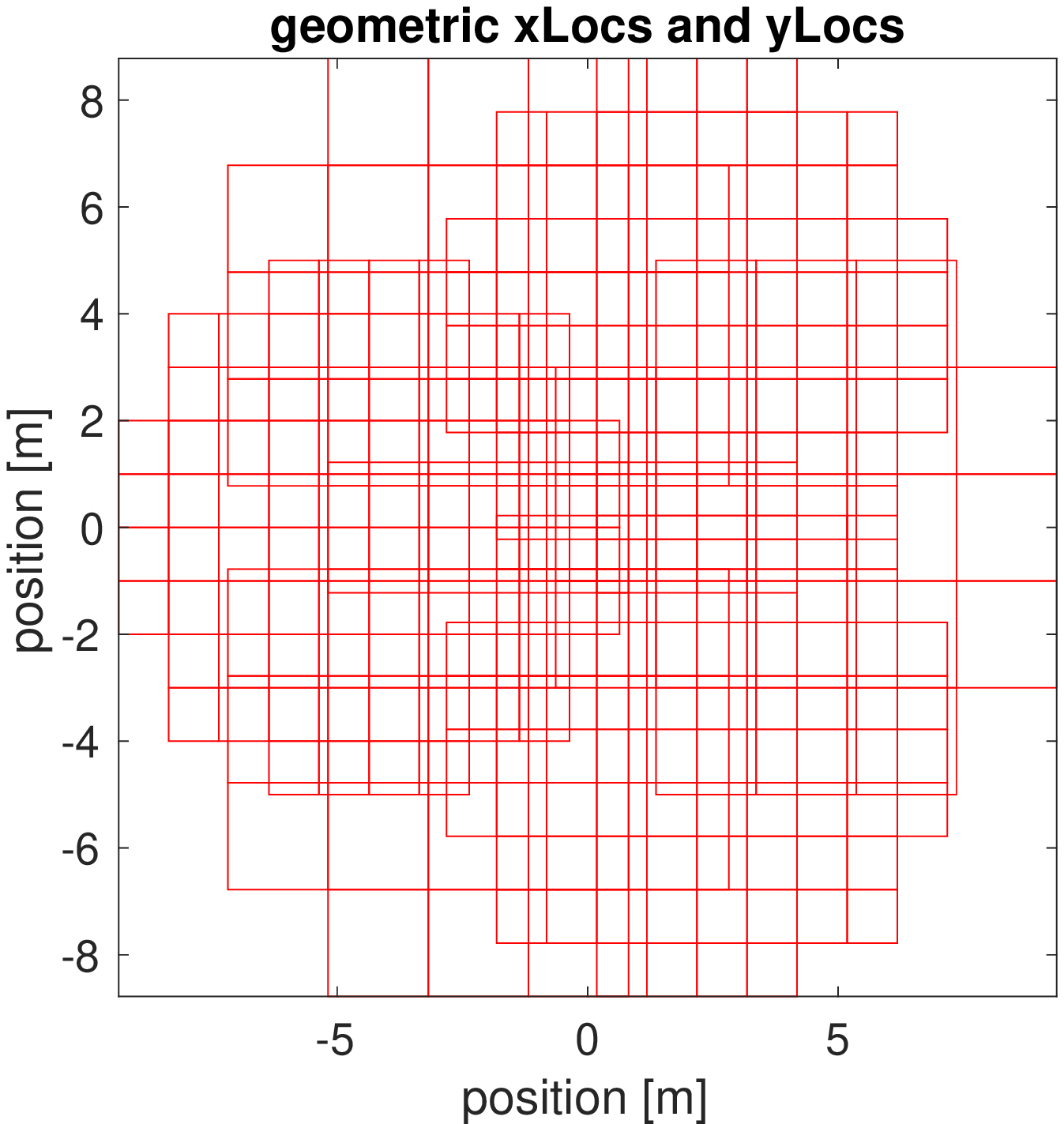}\includegraphics[width=0.35\textwidth,
             angle=0]{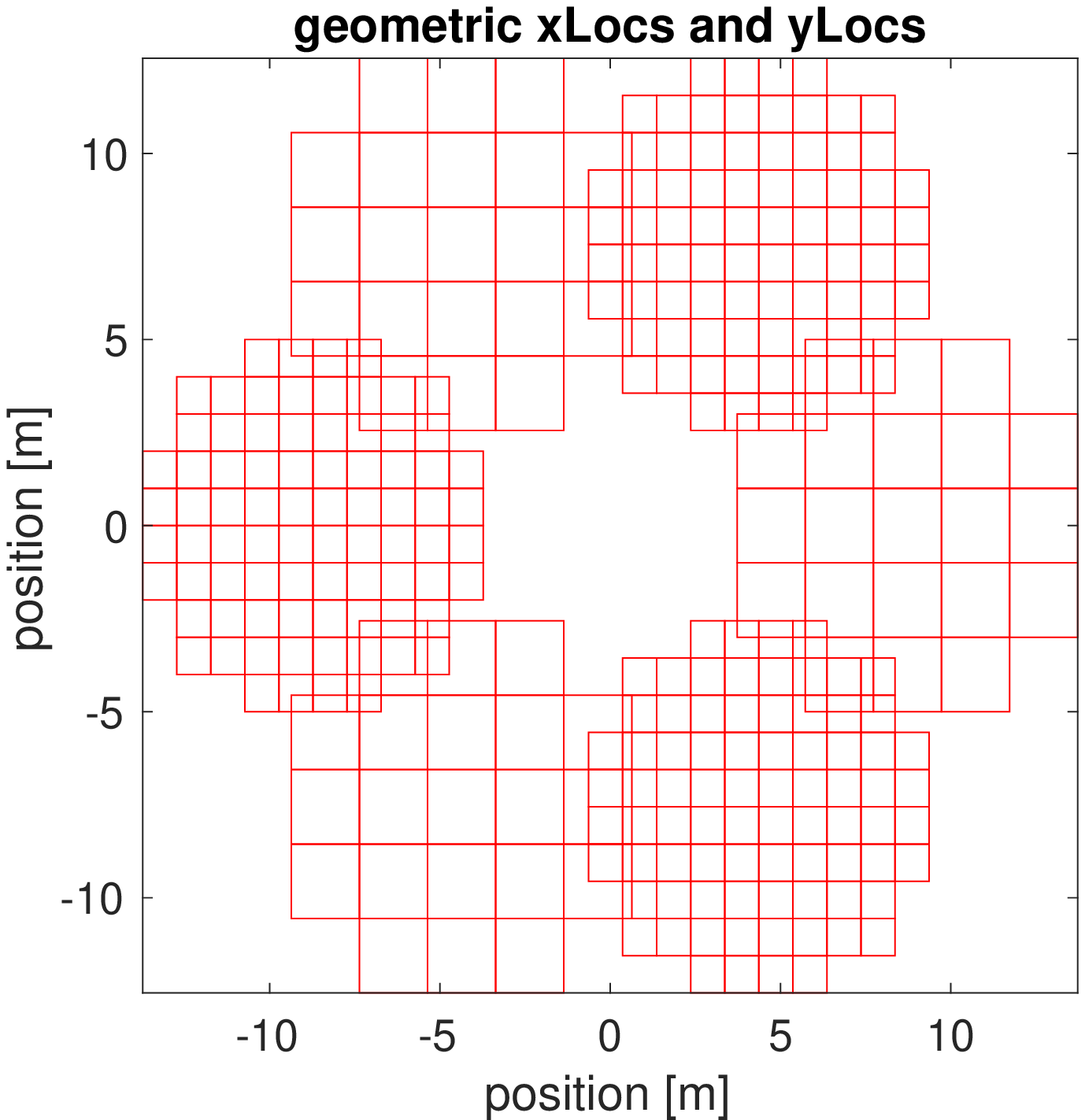}
	\end{center}
	\caption[]
	{\label{fig:hexagonalMixMeshGrids1}Six WFSs configuration with two kinds of subaperture, one being the double in size of the other, shown at three different conjugation altitudes creating growing separations.
 }
\end{figure*}

We have showed that both uniform and non uniform sampling schemes allow the reconstruction of a band-limited signal without aliasing, under some conditions presented in this paper. In addition to offsets and rotations, the AO design offers several other sampling schemes that facilitate SR. A non-exhaustive list is to: 
\begin{itemize}
    \item differently rotate and offset (possibly stretch) the WFS sampling arrays by non-integer multiples of each other (Fig. \ref{fig:hexagonalMeshRandRotGrids1}),
    \item use asymmetric guide star asterisms projected on-sky (Fig. \ref{fig:hexagonalMeshSpiralGrids1}),
    \item use WFSs with different numbers and sizes of subapertures (Fig. \ref{fig:hexagonalMixMeshGrids1}),
    \item allowing the subaperture sizes in each lenslet array to vary.
\end{itemize} 
In either case, the unique phase information collected by each WFS can be used to separate the aliased high-frequency from the low-frequency content of interest, and the higher-resolution wavefront can be accurately reconstructed.

\newpage

\end{appendix}
\end{document}